\documentclass[usenatbib]{mn2e}

\usepackage{graphicx} 
\usepackage{epsfig}

\def\Lya{Ly$\alpha$~} 
\def\OM{$\Omega_{\rm m}$~}

\def\LCDM{$\Lambda$CDM~} 

\def\OBh{$\Omega_{\rm b}h^{2}$~}

\def\HI{\hbox{H~$\rm \scriptstyle I\ $}} 
 
\def\HeI{\hbox{He~$\rm \scriptstyle I\ $}}
\def\HeII{\hbox{He~$\rm \scriptstyle II\ $}}

\def\etal{{\it et al.}~}

\title[Spatial fluctuations in the spectral shape of the UV background]{Spatial fluctuations
  in the spectral shape of the UV background at $2<z<3$ and the reionization of helium}

\author[J.S. Bolton \etal]
{James S. Bolton\thanks{E-mail:jsb@ast.cam.ac.uk}, Martin G. Haehnelt, Matteo Viel \& Robert F. Carswell\\
  Institute of Astronomy, University of Cambridge, Madingley
  Road, Cambridge, CB3 0HA\\}

\begin{document}

\date{1 December 2005}

\maketitle

\label{firstpage}

\begin{abstract}
  The low density hydrogen and helium in the IGM probed by QSO
  absorption lines is sensitive to the amplitude and spectral shape of
  the metagalactic UV background.  We use realistic \HI and \HeII \Lya
  forest spectra, constructed from state-of-the-art hydrodynamical
  simulations of a \LCDM Universe, to confirm the reliability of using
  line profile fitting techniques to infer the ratio of the
  metagalactic \HI and \HeII ionization rates.  We further show that
  the large spatial variations and the anti-correlation with \HI absorber
  density observed in the ratio of the measured \HeII to \HI column
  densities can be explained in a model where the \HI ionization rate
  is dominated by the combined UV emission from young star forming
  galaxies and QSOs and the \HeII ionization rate is dominated by
  emission from QSOs only.  In such a model the large fluctuations in
  the column density ratio are due to the small number of QSOs
  expected to contribute at any given point to the \HeII ionization
  rate.  A significant contribution to UV emission at the \HeII
  photoelectric edge from hot gas in galaxies and galaxy groups would
  decrease the expected fluctuations in the column density ratio.
  Consequently, this model appears difficult to reconcile with the
  large increase in \HeII opacity fluctuations towards higher
  redshift.  Our results further strengthen previous suggestions that
  observed \HeII \Lya forest spectra at $z\sim 2-3.5$ probe the tail
  end of the reionization of \HeII by QSOs.
\end{abstract}
 

\section{Introduction}

As primordial density fluctuations grow through gravitational Jeans
instability within CDM-like models, the web-like distribution of dark
matter strongly influences the clustering of baryonic matter on large
scales
(\citealt{Bi92,Cen92,Hernquist96,MiraldaEscude96,Theuns98,Zhang98}).
After the epoch of hydrogen reionization, the hydrogen gas present in
this baryonic component is highly ionized, leaving only a small
fraction of the gas as neutral atoms in photoionization equilibrium
within the intergalactic medium (IGM) .  This distribution of neutral
hydrogen is observable as the forest of \Lya absorption lines in the
spectra of high redshift QSOs (see \citealt{Rauch98} for a review).
Helium-4 is the second most abundant nuclide in the Universe after
hydrogen, with a mass fraction of $Y\simeq0.24$ ({\it e.g.}
\citealt{Walker91}).  Following the onset of \HeII reionization, the
structures responsible for the absorption seen in the \HI \Lya forest
should also be observable as discrete \HeII \Lya absorption lines at
the rest-frame wavelength of $304$ \AA~({\it e.g.}
\citealt{MiraldaEscudeOstriker90,MiraldaEscude93,MadauMeiksin94}).

The development of space-based UV-capable observatories in the 1990s
led to the first detection of a possible \HeII Gunn-Peterson trough in
the spectrum of $\rm Q0302-003$ by \cite{Jakobsen94} using the Faint
Object Camera on the {\it Hubble Space Telescope (HST)}.  Higher
resolution studies of the same spectrum using the Goddard High
Resolution Spectrograph and the Space Telescope Imaging Spectrograph
(STIS) on the {\it HST}, as well as spectral observations of $\rm HE$
$2347-4342$ using the {\it Far Ultraviolet Spectroscopic Explorer
  (FUSE)}, have all demonstrated a strong correlation between resolved
\HeII absorption features and those seen in the \HI \Lya forest,
confirming theoretical predictions.  Absorption from \HeII in the IGM
has so far been detected in six QSO spectra; $\rm PKS$ $1935-692$,
(\citealt{Tytler95,Jakobsen96,Anderson99}), $\rm HS$ $1700+64$,
(\citealt{Davidsen96}) $\rm HE$ $2347-4342$,
(\citealt{Reimers97,Kriss01,Smette02,Shull04,Zheng04b}), $\rm SDSS$
$\rm J2346-0016$, (\citealt{Zheng04a}), $\rm Q0302-003$
(\citealt{Jakobsen94,Hogan97,Heap00,Jakobsen03}) and most recently $\rm HS$
$1157-3143$, (\citealt{Reimers05a}).  The higher redshift \HeII
absorption measurements ($z \geq 2.8$) exhibit a patchy distribution,
with regions of high opacity interspersed by voids.  This observation,
coupled with the steady decline in the \HeII opacity at lower
redshift, along with measurements of the IGM temperature
(\citealt{Theuns02}), the evolution of the \HI \Lya forest effective
optical depth (\citealt{Bernardi03}) and measurements of the C~$\rm
\scriptstyle IV$/Si~$\rm \scriptstyle IV$ ratio
(\citealt{Songaila98}), has led to the suggestion that the final
stages of the \HeII reionization epoch may have occurred around $z
\simeq 3$.

In tandem with the observations, detailed models of the IGM have been
developed to examine the \HeII \Lya forest from a theoretical
perspective ({\it e.g.}
\citealt{ZhengDavidsen95,Bi97,Croft97,Zhang97,Theuns98,Wadsley98,Zheng98,Fardal98,Sokasian02,Gleser04}).
The \HeII \Lya forest is of particular interest as it is a sensitive
probe of the low density IGM (\citealt{Croft97}).  Additionally, the
ratio of \HeII to \HI column densities, $\eta = N_{\rm HeII}/N_{\rm
  HI}$, provides a constraint on the spectrum of the metagalactic UV
background (UVB) between $13.6$ $\rm eV$ and $54.4$ $\rm eV$.
Measurements of $\eta$ indicate a wide spread of values, from $\eta<1$
to $\eta>1000$ , implying that the metagalactic radiation field
exhibits inhomogeneities at the scale of $\simeq1$ Mpc
(\citealt{Reimers04,Shull04,Zheng04b}).  There is also some evidence
for an anti-correlation between $\eta$ and the density of the \HI \Lya
absorbers (\citealt{Reimers04,Shull04}).  Semi-analytical models
(\citealt{Heap00,Smette02}) suggest that a soft UVB with a significant
stellar contribution is required to reproduce the high opacity regions
($\eta>200$), while the observed opacity gaps ($\eta<100)$ have been
attributed to hard sources near the line-of-sight creating pockets of
highly ionized helium.  The small-scale variations in $\eta$ could
also be attributed to a spread in QSO spectral indices
(\citealt{Telfer02,Scott04}), local density variations
(\citealt{MiraldaEscude00}), finite QSO lifetimes
(\citealt{Reimers05b}), intrinsic absorption within the nuclei of
active galaxies (\citealt{Jakobsen03,Shull04}) or the filtering of QSO
radiation by radiative transfer effects (\citealt{Shull04,Maselli05}).
However, the exact nature of these fluctuations and their
interpretation is unclear.

In this paper we use realistic \HI and \HeII \Lya forest spectra,
constructed from state-of-the-art hydrodynamical simulations of a
\LCDM Universe, to test the reliability of using line profile fitting
techniques to infer the UVB softness parameter, defined as the ratio
of the \HI and \HeII metagalactic ionization rates, $S=\Gamma_{\rm
  HI}/\Gamma_{\rm HeII}$. We obtain improved estimates of the UVB
softness parameter and its uncertainty using published estimates of the \HI
and \HeII \Lya forest opacity. We concentrate on the redshift range
$2<z<3$ where the fluctuations of the mean \HeII opacity are still
moderate and \HeII reionization is probably mostly complete.  Finally,
we make a detailed comparison between the observed fluctuations in the
column density ratios of corresponding \HI and \HeII \Lya absorption
features measured by \cite{Zheng04b} (hereinafter Z04b) and those
expected due to fluctuations in a metagalactic \HeII ionization field
dominated by emission from QSOs and hot gas in collapsed haloes
(\citealt{Miniati04}) respectively.

The paper is structured as follows.  The method we use for obtaining
the UVB softness parameter from our simulations is described in
section $2$.  In section $3$ we present a simple model for UVB
fluctuations.  We discuss our line profile fitting method and compare
the UVB softness parameter inferred directly from our simulations to
the value obtained from the column density ratio $\eta$ in section
$4$.  In section $5$ we explore the effect of UVB fluctuations on the
measured column density ratio.  Our constraints on the UVB softness
parameter are given in section $6$, along with a discussion of
implications for the sources which contribute to the UVB.  We
summarise and conclude in section $7$.

\section{Simulations of the \HI and \HeII forest}
\subsection{Numerical simulations of the IGM}

\begin{table} 
  \centering
  \caption{Simulations used for our resolution and box size study.}
  \begin{tabular}{c|c|c|c|c|c|c|c}
    \hline
    Name     & Box size       & Particle & Mass resolution \\  
    & comoving Mpc   &  number & $h^{-1}M_{\odot}$/gas particle \\
  \hline
  15-400     & 15$h^{-1}$     & $2 \times 400^{3}$   & $6.78 \times 10^{5}$ \\
  15-200     & 15$h^{-1}$     & $2 \times 200^{3}$   & $5.42 \times 10^{6}$ \\
  30-400     & 30$h^{-1}$     & $2 \times 400^{3}$   & $5.42 \times 10^{6}$ \\
  15-100     & 15$h^{-1}$     & $2 \times 100^{3}$   & $4.34 \times 10^{7}$ \\
  30-200     & 30$h^{-1}$     & $2 \times 200^{3}$   & $4.34 \times 10^{7}$ \\
  60-400     & 60$h^{-1}$     & $2 \times 400^{3}$   & $4.34 \times 10^{7}$ \\
  30-100     & 30$h^{-1}$     & $2 \times 100^{3}$   & $3.47 \times 10^{8}$ \\
  \hline
  
  \label{tab:sims}
\end{tabular}
\end{table}

We have performed seven simulations of differing box size and mass
resolution using the parallel TreeSPH code GADGET-2
(\citealt{Springel01,Springel05}).  We shall only discuss the changes
made to the simulations for this work; unless otherwise stated all
other aspects of the code are as described in \cite{Bolton05}.  One
major difference is the incorporation of a non-equilibrium gas
chemistry solver for the radiative cooling implementation.  This
solves a set of coupled time-dependent differential equations which
govern the abundances of six species (H~$\rm \scriptstyle I$,
H~$\rm \scriptstyle II$, He~$\rm \scriptstyle I$, He~$\rm \scriptstyle
II$, He~$\rm \scriptstyle III$, $\rm e^{-}$) and the gas temperature
({\it e.g.}  \citealt{Cen92,Abel97,Anninos97,Theuns98}).  Previously
radiative cooling was implemented using the prescription of
\cite{Katz96}, which assumes ionization equilibrium.  This becomes a
poor approximation during \HI and \HeII reionization because gas
temperatures are underestimated; less atoms are ionized per unit time
compared to the non-equilibrium solution.  In addition, the radiative
recombination rates of \cite{Abel97} are adopted and their collisional
excitation cooling term for \HeI is added.  The collisional ionization
rates are taken from \cite{Theuns98}.

\begin{figure*}
  \centering
 \begin{minipage}{180mm}
   \begin{center}   
           
     \psfig{figure=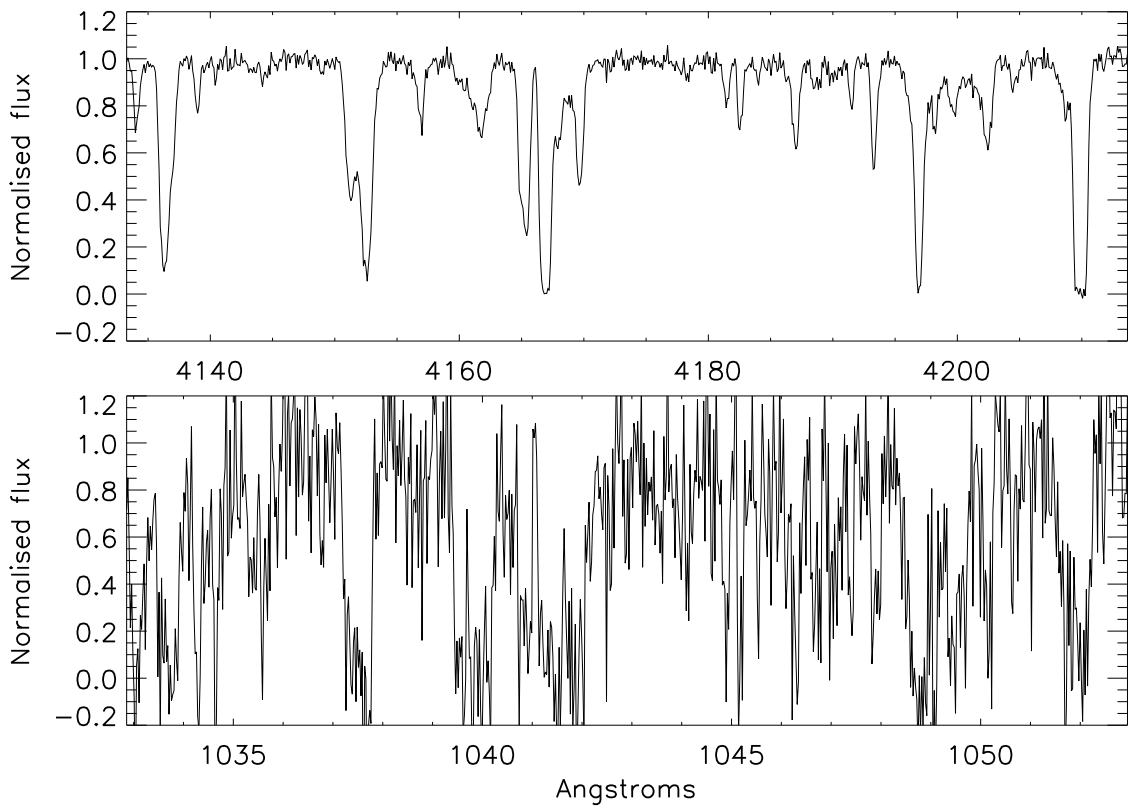,width=0.9\textwidth}
     \caption{
       Example of the artificial H~$\rm \scriptstyle I$ (top panel)
       and He~$\rm \scriptstyle II$ (bottom panel) \Lya forest spectra
       used for line profile fitting.  The spectra are constructed
       from the $z=2.4$ output of the $15-200$ model.  Each panel is
       constructed from four lines-of-sight. The resolution and noise
       properties resemble those of the observed spectra of Z04b.}
   \end{center}
   \label{fig:spec}
 \end{minipage}
\end{figure*}

For the UVB we have assumed a modified version of \cite{Madau99}
model, based on contributions from both QSOs and galaxies.  A redshift
mapping of the UVB was made so that \HI reionization begins at $z=6$
and \HeII reionization and reheating is postponed until $z \simeq
3.2$.  The photoionization balance is calculated assuming the gas is
optically thin, which leads to an underestimate of the photo-heating
rates (\citealt{AbelHaehnelt99,Bolton04}).  We have thus increased the \HeII
photo-heating rate to obtain gas temperatures consistent with
observations (\citealt{Ricotti00,Schaye00,McDonald01}).  The required
increase in the \HeII photo-heating rate was a factor $1.8$ at $z>
3.2$. This is somewhat smaller than in \cite{Bolton05} due to the
assumed late reionization of \HeII and the effect of the
non-equilibrium solver.

We adopt cosmological parameters consistent with those quoted by
\cite{Spergel03}, $(\Omega_{\rm m},\Omega_{\Lambda},\Omega_{\rm
  b}h^{2},h,\sigma_{8},n) = (0.26,0.74,0.024,0.72,0.90,1.0)$, and a
helium mass fraction of $Y=0.24$.  The simulations run for our
resolution study are listed in Table~\ref{tab:sims}.  In order to
investigate the dependence of the softness parameter on the assumed
cosmological model, four additional simulations were run with
$\Omega_{\rm m}=[0.22,0.30]$ and $\sigma_{8}=[0.8,1.0]$ using
$200^{3}$ gas and dark matter particles within a $15h^{-1}$ comoving
$\rm Mpc$ box.  All the other cosmological and astrophysical parameters
adopted for these additional simulations are the same as for the
resolution study.

\subsection{Artificial spectra and the H~$\rm \scriptstyle I$ and He~$\rm 
  \scriptstyle II$ effective optical depth} \label{sec:taueff}

For each simulation we have constructed artificial \HI and \HeII \Lya
forest spectra for $1024$ random lines-of-sight parallel to the
simulation box boundaries at eleven different redshifts.  The spectra
are generated at intervals of $\Delta z = 0.1$ within the redshift
range $2<z<3$.  Examples of the artificial spectra are shown in figure
$1$.  The pixel size and signal-to-noise ratio are chosen to mimic the
observed spectra used by Z04b in their analysis.  The \HI spectra are
binned onto pixels of width $0.1$~\AA~ and the signal-to-noise ratio
is $40$ per pixel.  The \HeII spectra have been binned onto pixels of
width $0.025$~\AA~ and the signal-to-noise ratio is $4$ per pixel.
Note that the pixel widths are identical in velocity space.

As described in more detail in the last section, the simulations have
been run using the updated UVB model of \cite{Madau99}.  As usual, we
have rescaled the \HI and \HeII optical depths in each pixel of the
simulated spectra to match the respective observed effective optical
depths, $\tau_{\rm eff} = -\rm ln \langle F \rangle$, where $\langle F
\rangle$ is the mean flux of the $1024$ lines-of-sight.  The \HI and
\HeII optical depths scale inversely with the \HI and \HeII ionization
rates; the metagalactic \HI and \HeII ionization rates can thus be
determined in this way ({\it e.g.}
\citealt{Rauch97,McDonaldMiraldaEscude01}).  We will use this later to
constrain the softness parameter of the UVB.

For the \HI effective optical depth the central values from the
fitting formula of \cite{Schaye03} have been used.  The corresponding
uncertainties have been estimated by binning the $1\sigma$ errors
given in table 5 of \cite{Schaye03} into redshift bins of width
$\Delta z = 0.1$.

Measurements of the \HeII effective optical depth, $\tau_{\rm HeII}$,
are more difficult to obtain; finding QSOs which have relatively clear
lines-of-sight and which are bright enough in the far-UV to enable the
detection of \HeII absorption spectra is challenging.  Only six QSO
spectra are known to show intergalactic \HeII absorption, and only
four have yielded \HeII opacity measurements.  It is therefore not
clear to what extent the available data provides a statistically
representative measure of the \HeII opacity.

\begin{figure}
\begin{center}
  \includegraphics[width=0.5\textwidth]{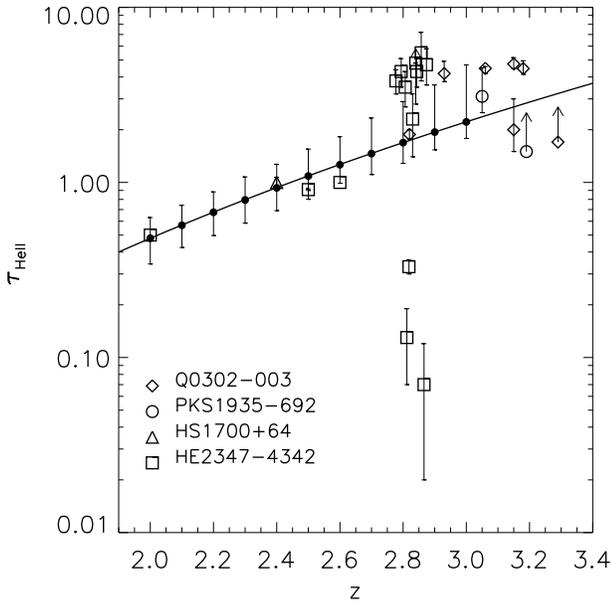}
\caption{
  Compilation of He~$\rm \scriptstyle II$ \Lya forest opacity
  measurements taken from the literature.  The solid curve is a fit to
  the data. The estimated uncertainty in this fit is shown by the
  filled circles with error bars.}
\label{fig:tau}
\end{center}
\end{figure}

Nevertheless, observational studies of the \HeII \Lya forest have met
with impressive success in recent years, in particular with high
resolution studies made using STIS and the {\it Far-Ultraviolet
  Spectroscopic Explorer (FUSE)}.  Figure~\ref{fig:tau} shows a
compilation of \HeII optical depth measurements taken from the
literature.  The data are from $\rm Q0302-003$
(\citealt{Jakobsen94,Hogan97,Heap00}), $\rm PKS$ $1935-692$
(\citealt{Jakobsen96,Anderson99}), $\rm HS$ $1700+64$
(\citealt{Davidsen96}) and $\rm HE$ $2347-4342$
(\citealt{Reimers97,Kriss01,Zheng04b}) with uncertainties shown where
applicable.  The solid curve in figure~\ref{fig:tau} shows the fit we
use as the mean \HeII effective optical depth for our artificial
spectra, given by:

\begin{equation}  \log \tau_{\rm HeII} = 0.345 + 5.324
\log\left(\frac{1+z}{4}\right). 
\label{eq:tau} \end{equation}

\noindent
Note that this fit only provides a general description for the global
evolution of the \HeII effective optical depth.  The \HeII \Lya forest
opacity exhibits strong fluctuations which increase rapidly with
increasing redshift; the metagalactic \HeII ionization rate is
obviously not spatially uniform (\citealt{Reimers04,Shull04,Zheng04b}).  The
error bars attached to the solid circles in figure~\ref{fig:tau}
provide an estimate of the variation in the \HeII opacity using a
simple model for a fluctuating UVB due to emission from QSOs.  The
error bars show the $25^{\rm th}$ and $75^{\rm th}$ percentiles of
$\tau_{\rm HeII}$ for each set of $1024$ spectra at the relevant
redshift.  We will turn to the details of modelling these UVB
fluctuations in section~\ref{sec:toymodel}.

\subsection{Numerical convergence}

As discussed by \cite{Theuns98} and \cite{Bolton05}, inferring the
metagalactic ionization rate by reproducing the observed effective
optical depth with artificial \Lya forest spectra stretches
present-day numerical capabilities due to the wide range of physical
scales involved. It is thus important to perform numerical convergence
tests.  We have used the simulations listed in Table~\ref{tab:sims}
for a mass resolution and box size study.  The $400^{3}$ simulations
are only run to $z=3$; they were too computationally demanding to
enable a practical study below this redshift.  However, the $z=3$ data
should provide a good indication of how close the simulations are to
convergence.

As box size is increased, larger voids can be accommodated within the
simulation, reducing the mean \Lya absorption and altering the \HI and
\HeII gas distribution in a similar manner.  There is an $8$ per cent
reduction in both the \HI and \HeII ionization rates inferred from the
$30-200$ model compared to the $15-100$ run with the same mass
resolution, with a further $3$ per cent reduction for the $60-400$
data.  Note that the ratio of the inferred ionization rates for the
two species depends little on box size.

As noted by \cite{Theuns98}, convergence of the \HeII opacity requires
higher numerical resolution than the \HI opacity.  Increasing the mass
resolution of a simulation will resolve smaller haloes, transferring
more optically thin gas from the low density IGM into optically thick,
high density regions, decreasing the mean \Lya absorption.  The
reduction of the \HI effective optical depth resulting from this
re-distribution of the low density gas is partially offset by the
increased contribution to the opacity from the high density regions.
For He~$\rm \scriptstyle II$, which is optically thick down to much
lower gas densities, this offset is less pronounced, producing a
greater overall change in the \HeII effective optical depth for
increased mass resolution.  The \HI ionization rate inferred from the
$15-200$ model is $8$ per cent lower compared to the value of the
$15-100$ run, with a further drop of $8$ per cent for the $15-400$
model. The HI opacity has only marginally converged at $z=3$.  As
expected the \HeII ionization rate shows a greater reduction in
magnitude with increasing mass resolution, exhibiting a $20$ per cent
reduction between the $15-200$ and $15-400$ models and a $33$ per cent
drop for the $30-200$ and $30-400$ simulations.  Note that the
stricter resolution requirement for $\tau_{\rm HeII}$ compared to
$\tau_{\rm HI}$ has implications for the softness parameter we will
infer later.  For lower mass resolution the inferred softness
parameter will be smaller compared to that obtained from identical
higher resolution simulations.

We estimate that the ratio of the \HeII and \HI ionization rates
inferred from the $15-200$ simulation will be systematically low by at
least $12$ per cent in the redshift range $2<z<3$.  We shall use this
to correct our final estimation of $S$ in section~\ref{sec:UVB}.

\section{A simple model for UVB fluctuations} \label{sec:toymodel}
\subsection{Fluctuations in the \HeII opacity due to QSOs}

At $2<z<3$ fluctuations in the \HI metagalactic ionization rate are
expected to be small ({\it e.g.}
\citealt{Croft99,MeiksinWhite04,Croft04}) and the observed opacity
variation is generally believed to be due to density fluctuations in
the IGM.  However, the spatial fluctuations in the \HeII opacity are
significantly larger than those of the \HI opacity at the same
redshift. Furthermore, the ratio of the \HeII to \HI opacity also
shows large spatial fluctuations
(\citealt{Reimers04,Shull04,Zheng04b}).  The only plausible explanation
for these observations are substantial fluctuations in the
metagalactic radiation field at the \HeII photoelectric edge. If, as
is generally assumed, the \HeII ionizing flux is produced by QSOs this
conclusion is not too surprising.  The comoving attenuation length for
\HeII ionizing photons is about a factor of ten smaller than that for
\HI ionizing photons (\citealt{Fardal98,MiraldaEscude00}) and thus is
similar or shorter than the mean distance between bright QSOs. As
discussed by \cite{Fardal98} this is expected to result in substantial
fluctuations in the metagalactic \HeII ionization rate.  The details
of these fluctuations will depend on the luminosity function, life
time, spectral energy and angular distribution of the emission of QSOs
as well on details of the spatial distribution of He~$\rm \scriptstyle
II$.  A full numerical model of this problem including self-consistent
radiative transfer around point sources is probably beyond present-day
numerical capabilities (see \citealt{Sokasian02,Maselli05} for
attempts to implement some of the radiative transfer effects)
and certainly beyond the scope of this paper.  We will instead use a
simple ``toy model'' similar to that employed by \cite{Fardal98} which
should incorporate many if not most of the relevant aspects of a
fluctuating \HeII ionization rate.
 
We first assume that at any given point all QSOs within a sphere of
radius equal to the typical attenuation length, $R_{0}$, contribute to
the \HeII ionization rate.  This assumption will approximate the
effect of an optically thick medium on the propagation of ionizing
radiation.  Note however that a fully self-consistent treatment of
radiative transfer will be required to accurately model this process.
The \HeII ionization rate at redshift $z$ can then be written as,

\begin{equation} \Gamma_{\rm HeII}(z) = \sum_{j=1}^{N_{\rm QSO}} 
\left[ \int^{\infty}_{\nu_{228}} \frac{L_{\rm j}(\nu)}{4 \pi
R_{j}^{2}} 
\frac{\sigma_{\rm HeII}(\nu)}{h \nu} \hspace{1mm}  d\nu \right],  
\label{eq:flucs} \end{equation}

\noindent
for $R_{\rm j} < R_{0}/(1+z)$, where $R_{\rm j}$ is the proper
distance of QSO $\rm j$ with luminosity $L_{j}(\nu)$, $\sigma_{\rm
  HeII}(\nu)$ is the \HeII photoionization cross-section and all other
symbols have their usual meaning.  We then populate collapsed haloes
in our simulation with QSOs by Monte-Carlo sampling the observed QSO
luminosity function.  The comoving attenuation length of \HeII
ionizing photons we use is,

\begin{equation} R_{0} = 30 \left( \frac{1+z}{4} \right)^{-3}
\hspace{1mm} \rm Mpc, \end{equation}

\noindent
assuming the number of Lyman limit systems per unit redshift is
proportional to $(1+z)^{1.5}$ (\citealt{StorrieLombardi94}) and using
a normalisation based on the model of \cite{MiraldaEscude00}.  Note
that this length is somewhat larger than our simulation box.  We have
thus tiled identical copies of our periodic 15-200 simulation box in
all three spatial directions around a central ``master box'' to create
a volume large enough to fully contain the ``attenuation sphere''
around each point in the master box from which the artificial spectra
are produced.

We have used the fit to the optical luminosity function (OLF) of QSOs
obtained by \cite{Boyle00} in the $\rm b_{J}$ band (effective
wavelength 4580\AA),

\begin{equation} \phi(L,z) = \frac{\phi_{*}/L_{*}(z)}{[L/L_{*}(z)]^{-\beta_{1}} + 
  [L/L_{*}(z)]^{-\beta_{2}}}. \end{equation}

\noindent
For the evolution of the break luminosity, $L_{*}(z)$ we have taken
the pure luminosity evolution model of \cite{Madau99},

\begin{equation} L_{*}(z) = L_{*}(0)(1+z)^{\alpha_{s} -1} \frac{ e^{\zeta z} (1  +
  e^{\xi z_{*}})} { e^{\xi z} + e^{\xi z_{*}} }. \end{equation}

\noindent
The value of the model/fit parameters are $\zeta=2.58$, $\xi=3.16$,
$z_{*}=1.9$, $\beta_{1}=-3.41$, $\beta_{2}=-1.58$, $\rm
log(\phi_{*}/Gpc^{-3}) = 2.99$ and $\rm log(L_{*}(0)/L_{\odot})=10.67
$ for $\Omega_{\rm m}=0.3$, $\Omega_{\Lambda}=0.7$ and $h=0.7$.  A
power law spectral distribution for the QSO luminosity, $L(\nu)
\propto \nu^{-\alpha_{s}}$, is assumed.  Note that more recent
estimates of the OLF exist, ({\it e.g.} \citealt{Croom04,Richards05}),
but we adopt the above as it is the basis of the updated
\cite{Madau99} UVB model to which we compare our data.

Collapsed haloes within the simulation volume have been identified
using a friends-of-friends halo finding algorithm with a linking
length of 0.2.  We have based the assignment of a particular
QSO luminosity to a collapsed halo on the empirically determined
relation between black hole mass and the velocity dispersion of its
host bulge (\citealt{FerrareseMerritt00}).  For a QSO luminosity
$L_{\rm QSO}$ randomly drawn from the OLF, the velocity dispersion of
the host bulge is approximately:

\begin{equation} \sigma \simeq 200 \rm \hspace{1mm} 
kms^{-1} \left( \frac{L_{\rm QSO}}{4.61\times 10^{12} \hspace{1mm} 
\epsilon \hspace{1mm} L_{\odot}} \right)^{0.21}  \end{equation}

\noindent
where $\epsilon$ is the radiative efficiency of the black hole as a
fraction of the Eddington limit.  We adopt $\epsilon = 0.1$.  The halo
with the velocity dispersion closest to this value is then chosen and
its position is determined by randomly selecting the identified halo
within one of the boxes tiled to create the total simulation volume.
The total number of QSOs brighter than absolute magnitude $M_{\rm
  lim}$ in our comoving simulation volume, $V$, at redshift $z$ is
then required to satisfy:

\begin{equation} N_{\rm QSO}(z,M_{\rm b_{\rm J}} < M_{\rm lim}) = 
V \int^{M_{\rm lim}}_{-\infty} \phi(M_{\rm b_{\rm J}},z) \hspace{1mm}
dM_{\rm b_{\rm J}} \end{equation}

\noindent
where we adopt $M_{\rm lim} = -22$.

Following \cite{Madau99}, we adopt a broken power law for the QSO
spectral energy distribution:

\begin{equation}
L({\nu})\propto \cases{\nu^{-0.3} &($2500<\lambda<4600\,$\AA),\cr
  \noalign{\vskip3pt}\nu^{-0.8} &($1050<\lambda<2500\,$\AA),\cr
  \noalign{\vskip3pt}\nu^{-\alpha} &($\lambda<1050\,$\AA).\cr}
\end{equation}

\noindent
The extreme-ultraviolet (EUV) spectral index, $\alpha$, is a variable
parameter in our model.  Each QSO placed in the simulation volume has
a value for $\alpha$ assigned to it by Monte Carlo sampling the
distribution of QSO spectral indices in the EUV.  The distribution is
based on the data of \cite{Telfer02} for radio quiet QSOs, taken to be
a Gaussian with mean $1.61$ and standard deviation $0.86$.  Note,
however, lower redshift {\it FUSE} observations (\citealt{Scott04})
favour a harder EUV spectral index.

When constructing artificial spectra using our fluctuating UVB model,
the opacity in each pixel in the artificial \HeII spectra is rescaled
linearly with the new fluctuating ionization rate computed using
equation~\ref{eq:flucs}.  The new set of spectra are then rescaled
once more to match the observed \HeII effective optical depth.  We do
not alter the opacity of the corresponding \HI spectra; a uniform UVB
at the hydrogen photoelectric edge should be a reasonable
approximation at the redshifts we consider ({\it e.g}
\citealt{Croft99,MeiksinWhite04,Croft04}).  Note that we have also
used this fluctuating UVB model to estimate the uncertainty in the
\HeII effective optical depth in figure~\ref{fig:tau}.

\subsection{Fluctuations in the \HeII opacity due to thermal emission
  from hot gas}

It has recently been suggested by \cite{Miniati04} (M04 hereinafter)
that thermal radiation from shocked gas in collapsing haloes may
provide a substantial contribution to the UVB at the \HeII
photoelectric edge.  M04 show that, for plausible assumptions for the
distribution of hot gas, bremsstrahlung and line emission could
provide a similar number of \HeII ionizing photons compared to QSOs at
$z=3$, and may even dominate the total photon budget above the \HeII
photoelectric edge for $z>4$.  However, within the average attenuation
length for an \HeII ionizing photon, there will be many more haloes
contributing to the UVB via thermal emission compared to QSOs, which
have a rather small duty cycle.  It is therefore unclear whether an
\HeII metagalactic ionization rate dominated by emission from hot gas
will reproduce the observed fluctuations in the \HeII opacity, or
whether the more uniform distribution of sources expected for thermal
emission will damp fluctuations in the metagalactic radiation field at
the relevant physical scale.

We modify our existing model for a fluctuating UVB due to QSOs to
explore this.  Following M04, a halo of mass $M$ will have virial
temperature,

\begin{equation}  T(M) \simeq 2.3 \times 10^{6} \left(\frac{M}{10^{12}M_{\odot}}\right)^{2/3}\left(\frac{1+z}{4}\right) \hspace{2mm} \rm K. \end{equation}

\noindent
The number of haloes emitting ionizing radiation at a given redshift
will then be constrained by the radiative lifetime of the gas in each
halo.  Using the prescription of M04, the ratio of the gas cooling
time to the Hubble time for a halo of mass $M$ is given by,

\[  \frac{t_{\rm cool}}{t_{\rm Hubble}} \simeq 0.21 \left(\frac{M}{10^{12}M_{\odot}}\right)^{2/3}\left(\frac{1+z}{4}\right)^{-1/2} \]
\begin{equation} \hspace{15mm} \times \left[ 1 + 0.45 \left(\frac{T(M)}{10^{6} \hspace{1mm} \rm K}\right)^{-1}\right], \label{eq:cool} \end{equation}

\noindent
where the last term takes into account the expected extension to the
gas cooling time due to energy injection by supernovae.  We use
equation~\ref{eq:cool} to determine the probability that each halo
within the simulation volume is radiating, thus providing a lower
limit to the number sources within an attenuation sphere at any given
redshift.  We assume the monochromatic luminosity of the hot gas in
the halo scales as $MT(M)^{-1/2}$, and the spectrum of the
bremsstrahlung emission is flat ($f_{\nu} = \mathrm{constant}$) for
the purposes of computing $\Gamma_{\rm HeII}$.  We then adjust the
flux of the radiating haloes with a global factor such that the
integrated flux reproduces the observed \HeII optical depth. This
assumption is extreme in the sense that it leaves no room for the
contribution by QSOs.  Note, however, that our assumption for the mean
free path of \HeII ionizing photons is probably an underestimate for
the rather hard bremsstrahlung spectrum.

\section{Measuring the softness parameter from \HI and \HeII \Lya forest absorption lines}

\subsection{The softness parameter}

A combined analysis of the \HI and \HeII forest makes it possible to
constrain the spectral shape of the UVB. This is normally done by
defining the ratio of the metagalactic \HI and \HeII ionization rates
as a softness parameter,

\begin{equation} S = \frac{\Gamma_{\rm HI}}{\Gamma_{\rm HeII}}, \end{equation}

\noindent
This quantity provides a direct constraint on the spectral shape of
the UVB between the \HI and \HeII photoelectric edges at $912$~\AA~
and $228$~\AA.  Assuming the hydrogen and helium gas in the IGM is in
ionization equilibrium, the ratio of observed column densities $N_{\rm
  HeII}/N_{\rm HI}$ measured from absorption lines in the \HI and
\HeII \Lya forest can be related to the softness parameter by ({\it
  e.g.} \citealt{MiraldaEscude93,Fardal98})

\begin{equation} \eta = \frac{N_{\rm HeII}}{N_{\rm HI}} = S 
\frac{n_{\rm HeIII}}{n_{\rm HII}} \frac{\alpha_{\rm
HeIII}}{\alpha_{\rm HII}}, \label{eq:etaeq} 
\end{equation}
 
\noindent 
where $n_{\rm i}$ and $\alpha_{\rm i}$ are the number density and
radiative recombination coefficient for species $i$.  If the IGM is
highly ionized,

\begin{equation} \frac{n_{\rm HeIII}}{n_{\rm HII}} \simeq \frac{n_{\rm He}}{n_{\rm H}} = \frac{Y}{4(1-Y)}. \label{eq:Hefrac} \end{equation}
 
\noindent
Adopting a helium mass fraction of $Y=0.24$ and evaluating the
radiative recombination coefficients of \cite{Abel97} at $T=15000$ $\rm
K$:

\begin{equation} S \simeq  2.4 \eta. \label{eq:soft} \end{equation}

\noindent
It is important to note that equation~\ref{eq:soft} is an
approximation and as such will not provide an exact conversion from
the observable $\eta$ to $S$.  In particular, if \HeII is still
undergoing reionization equation~\ref{eq:soft} will underestimate the
true softness parameter; not all the helium in the IGM will be doubly
ionized and equation~\ref{eq:Hefrac} will no longer hold.  Similarly,
if \HeII is still being reionized, the assumption of ionization
equilibrium will no longer hold and equation~\ref{eq:soft} will
overestimate the true softness parameter.  There is also a weak
temperature dependence in this relation depending on the assumed
values for the recombination coefficients ({\it e.g.}
\citealt{Shull04}).  However, we expect this approximation is accurate
to within $5$ per cent for the redshift range we consider.
Equation~\ref{eq:soft} will be adopted throughout this paper to
convert from $\eta$ to $S$.

\subsection{Fitting line profiles to the \HI and \HeII \Lya forest}

\begin{figure*}
  \centering
  \begin{minipage}{180mm}
    \begin{center}
      \psfig{figure=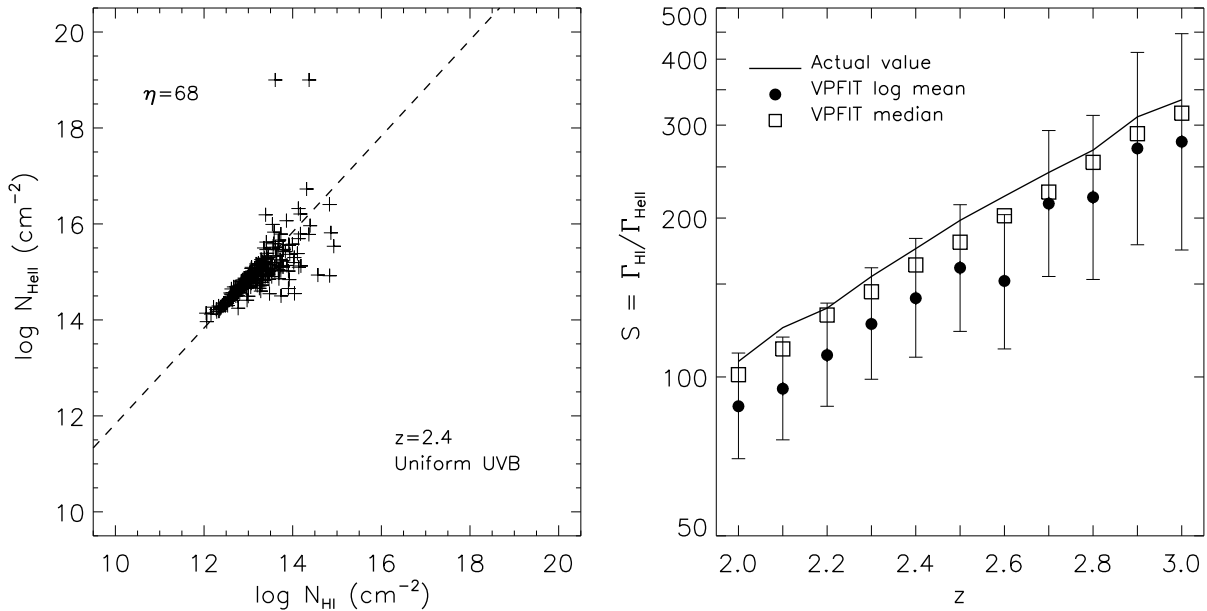,width=1.0\textwidth}
      \caption{
        {\it Left:} Scatter plot of the \HeII against \HI column
        density obtained by fitting Voigt profiles to artificial \Lya
        forest spectra at $z=2.4$.  A uniform UVB has been assumed.
        The dashed line indicates the median column density ratio of
        the sample.  {\it Right:} Comparison of the softness
        parameter, $S$, obtained by rescaling the artificial \Lya
        forest spectra to match the observed \HI and \HeII effective
        optical depths (solid line) with the value determined by
        fitting Voigt profiles to the absorption features.  The filled
        circles with error bars show the logarithmic average of $S$
        and the associated uncertainty, as obtained from the column
        density ratio $\eta$ using equation~\ref{eq:soft}.  The open
        squares indicate the median value of $S$. }
      \label{fig:compare}
    \end{center}
  \end{minipage}
\end{figure*}

In their recent study of the helium reionization history, Z04b present
\HI and \HeII \Lya forest spectra of the quasar $\rm HE$ $2347-4342$
obtained using the Very Large Telescope (VLT) and {\it FUSE}.  They
analyse the spectra by fitting Gaussian profiles to the \HI lines in
the VLT spectrum.  Once the \HI components are identified, counterpart
\HeII lines are defined in the {\it FUSE} spectrum, constrained to
have the same redshift and line width as the \HI lines; only the
column density is allowed to vary.  In the event that \HeII optical
depth is too high to produce reasonable values of $\eta$, additional
\HeII sub-components are added with a line width $b_{\rm HeII} = 27$ $\rm
kms^{-1}$.  The assumption of equal line widths is based on the
analysis of a sub-sample of unblended \HI and \HeII lines by Z04b;
they find a line width ratio of $\xi = b_{\rm HeII}/b_{\rm HI} = 0.95 \pm 0.12$.  In this
way Z04b have estimated the column density ratio in the redshift range
$2.0<z<2.9$.  Z04b detect large variations in the column density ratio
on small scales ($\Delta z \simeq 0.001$) and have interpreted these
as spatial fluctuations in the shape of the UVB.  \cite{Shull04} who
used an independent pixel-by-pixel optical depth technique to analyse
the same observational data come to similar conclusions.  However, as
pointed out by \cite{FechnerReimers04} and \cite{Shull04}, values of
$\eta<10$ and $\eta>400$ should be considered with some caution and
are probably not real. They are likely to be caused by uncertainties
introduced by background subtraction, the low signal-to-noise of the
spectra, line saturation and blending with higher order Lyman series
lines and metals.

We have performed the same analysis as Z04b on our artificial \Lya spectra.
This enables us to assess if there is a substantial contribution from
errors in the line fitting method to the observed variations in $\eta$
and also to determine how well the softness parameter can be
recovered.  We have analysed our artificial spectra using an automated
version of the Voigt profile fitting package VPFIT (Carswell {\it et
  al.}, http://www.ast.cam.ac.uk/$\sim$rfc/vpfit.html), modified to
simultaneously fit \HI and \HeII absorption lines.  The automated
Voigt profile fitting procedure is run on $50$ random lines of sight
taken from outputs produced by the $15-200$ simulation.  For each \HI
line which is identified, a corresponding \HeII line is fitted at the
same redshift.  Additional \HeII lines are added where no
corresponding \HI can be identified.  Turbulent line broadening is
assumed, such that the ratio of line widths is fixed to be $\xi =
b_{\rm HeII}/b_{\rm HI} = 1.0$, consistent with the Z04b result.  This
should be a reasonable assumption; Hubble broadening is expected to be
the dominant contribution to the width of absorption lines in the \Lya
forest (\citealt{Weinberg98}).  Note, however, that additional motions
in the IGM velocity field from convergent flows and shocks may produce
a departure from pure turbulent broadening for some of the lines.  The
maximum column density allowed in the fitting procedure is fixed at
$10^{19}$ $\rm cm^{-2}$.  Similarly, the upper limit to the width of
the fitted line profiles is set at $80$ $\rm kms^{-1}$.

\subsection{Measuring the softness parameter using line profile fitting}
\begin{figure*}
  \centering
  \begin{minipage}{180mm}
    \begin{center}
      \psfig{figure=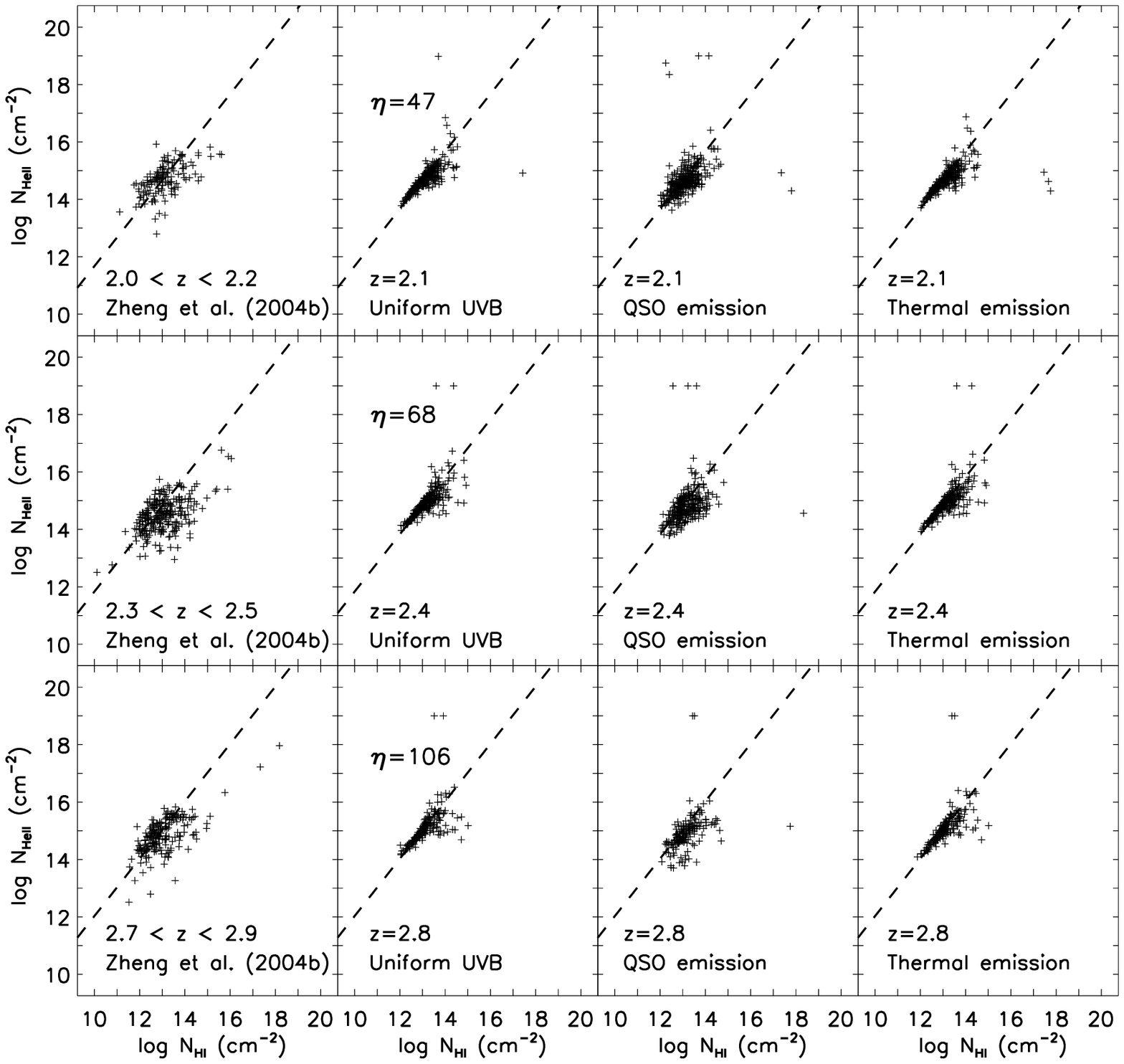,width=1.0\textwidth}
    \caption{
      Scatter plots of the column density ratio $\eta$ measured by
      Z04b (left column), from artificial spectra constructed with a
      uniform UVB at the \HeII photoelectric edge (second column), a
      fluctuating UVB due to QSOs (third column) and a UVB dominated
      by emission from hot gas in collapsing haloes (right column).
      The redshift ranges are indicated on each panel.  The dashed
      line shows the median column density ratio of the artificial
      spectra produced with a uniform UVB, shown in the second column.
      To facilitate comparison, the same line is repeated in the other
      panels.}
    \label{fig:eta}
  \end{center}
\end{minipage}
\end{figure*}

The left panel of figure~\ref{fig:compare} shows a scatter plot of the
column density ratio $\eta$ we have obtained from our artificial
spectra at $z=2.4$. A uniform background was assumed.  When
determining $\eta$ we only use line pairs which have well defined
profile fits; any line with an error in the column density greater
than a factor of two is rejected.  The \HI and \HeII column densities
are tightly correlated with a moderate scatter.

In the right panel of figure~\ref{fig:compare} we compare the softness
parameter calculated from the column density ratio using
equation~\ref{eq:soft} to the actual value used to produce the
artificial spectra at $2<z<3$.  Again we assume a spatially uniform
UVB.  For the range of values under consideration ($100<S<350$), line
fitting recovers the actual value of the softness parameter well. If
we take the median column density ratio (open squares) the recovered
value is on average $7$ per cent less than the actual value used to
produce the artificial spectra.  Following Z04b, we also plot the
logarithmic mean of the softness parameter $\log(\bar{S}) =
[\sum\log(S_{\rm i})]/n$ (filled circles with error bars).  The
uncertainty on the logarithmic mean, $\log(\Delta \bar{S}) =
[\sum[\log(\Delta S_{\rm i})]^{2}]^{1/2}/n$, is computed by
propagating the uncertainties on the \HI and \HeII column densities.
The median appears to provide the better measure of the softness
parameter; it is not weighted so heavily in favour of the lower column
density lines which are more commonly identified by the fitting
package.  As might be expected, we find fitting Voigt profiles to only
\HI and \HeII \Lya absorption features will not recover the softness
parameter accurately once the \HeII lines are predominantly saturated.
In this case, only higher order \HeII absorption lines or an accurate
determination of the \HeII effective optical depth from a large
statistical sample of \HeII absorption spectra would enable the
recovery of $S$.

Overall, we conclude that the line fitting method of Z04b will recover
the spectral shape of the UVB along the line-of-sight they consider
well, although one should keep in mind that the number of observed
spectra is very small, and systematic uncertainties in the line
fitting procedure means one should view the extreme values of $\eta$
with a certain amount of scepticism
(\citealt{FechnerReimers04,Shull04}).  It is also clear that the
magnitude of the observed scatter in the column density ratio is much
larger than our results for a spatially homogeneous UV background. We
will discuss this in more detail in the next section.

\section{Modelling the observed fluctuations in the column density ratio} \label{sec:obsfluc}
\subsection{The magnitude and physical scale of the fluctuations}

The column density ratio, $\eta=N_{\rm HeII}/N_{\rm HI}$, measured by
Z04b exhibits large variations ($1<\eta<1000$) and the evolution of
the median softness parameter is consistent with a UVB with a
relatively hard spectrum, $\eta<200$. It has been suggested that
spatial fluctuations in the UVB at the \HeII photoelectric edge could
produce the observed \HeII opacity variations ({\it e.g.}
\citealt{Reimers04,Shull04,Zheng04b}).  So far we have adopted a
spatially uniform UVB in the optically thin limit. At the \HI
photoelectric edge for $2<z<3$ the typical mean free path of an
ionizing photon should be much longer than the distance between
ionizing sources; a large number of sources are expected to contribute
to the \HI ionization rate at any given point.  Hence, as discussed
previously, the observed variations in $\eta$ must be caused by
fluctuations in the UVB at the \HeII photoelectric edge.  We will now
use the model described in section~\ref{sec:toymodel} to investigate
whether such fluctuations can explain the distribution in $\eta$
observed by Z04b.

\begin{figure*}
  \centering
  \begin{minipage}{180mm}
    \begin{center}
      \psfig{figure=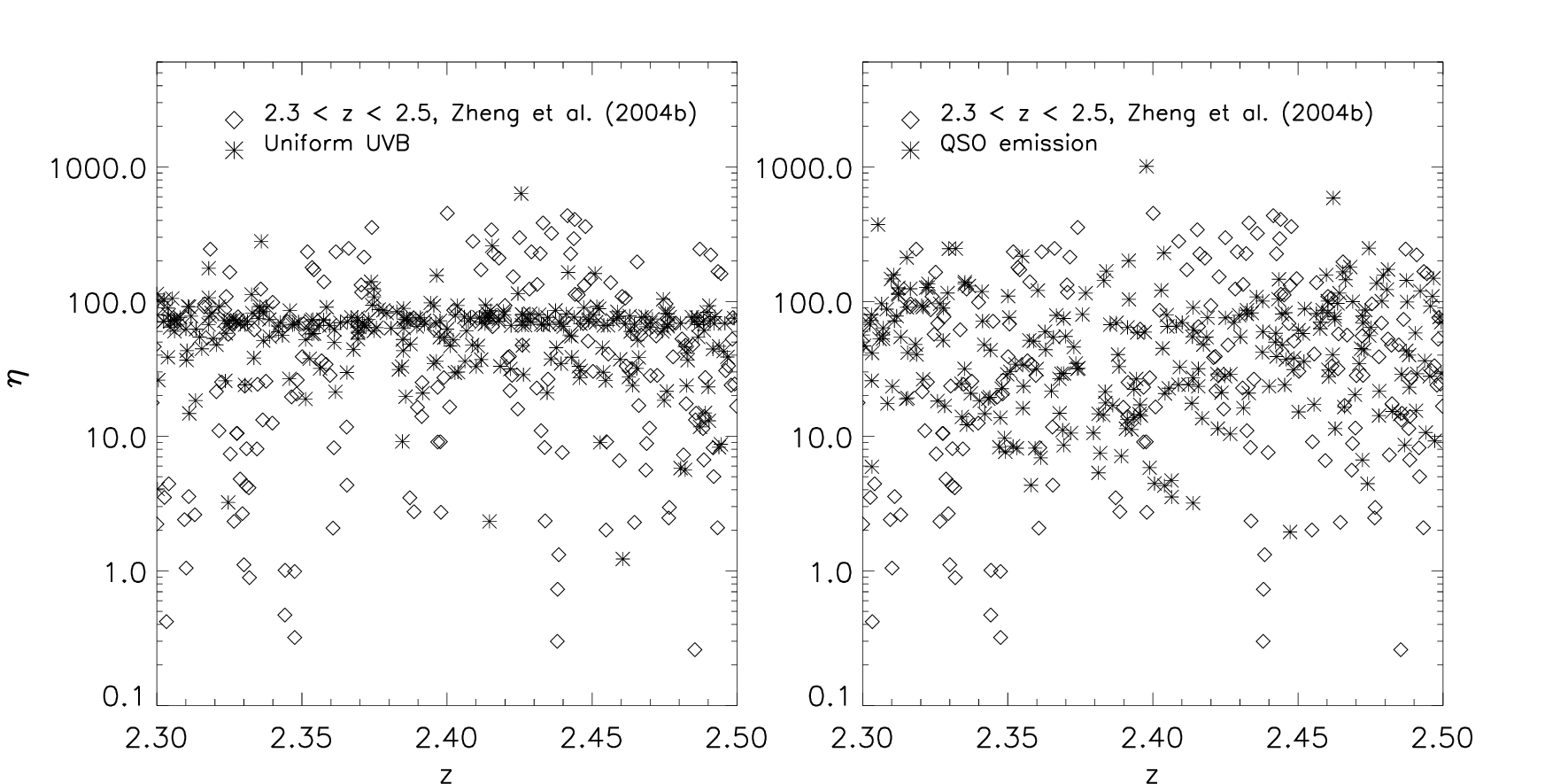,width=1.0\textwidth}
      \caption{
        Scatter plots of the column density ratio against redshift for
        $2.3 < z < 2.5$.  The observational data of Z04b are
        represented by open diamonds in both panels.  In the left
        panel, the column density ratios determined from our
        artificial spectra, generated using a uniform UVB at $z=2.4$,
        are indicated by stars. The redshifts for the simulated
        spectra have been shifted such that they cover the same
        redshift range as the observational data while maintaining the
        physical separation between data points. The actual simulation
        output is at at $z=2.4$.  The stars in the right panel show
        the column density ratios obtained from artificial spectra
        generated with a fluctuating UVB due to QSOs.  }
      \label{fig:scatter}
    \end{center}
  \end{minipage}
\end{figure*}

The left column of figure~\ref{fig:eta} shows the column density ratio
$\eta$ at $z=[2.1,2.4,2.8]$ as measured by Z04b. The second column
shows the same measured from our artificial spectra assuming a
spatially uniform background.  The scatter in $\eta$ for our model
spectra is clearly significantly smaller than observed.  The observed
scatter is thus unlikely to be explained by uncertainties in
identifying the correspondence between \HI and \HeII absorption lines
or measuring their column densities. The third column of the plot
shows $\eta$ obtained from spectra constructed using our model for
spatial fluctuations in the \HeII ionization rate due to QSOs.  The
scatter is significantly larger compared to the uniform UVB case,
suggesting that UVB fluctuations do indeed play an important role in
producing the observed variation in the \HeII opacity and softness
parameter.  The total scatter is around $2.0$ dex, about $0.5$ dex
smaller than that of the Z04b data.  It is not obvious how significant
this rather small difference is; Voigt profile fitting is a somewhat
subjective procedure and we have been rather strict in omitting bad
fits.  Nevertheless, it may indicate that we have over-estimated the
attenuation length, under-estimated the scatter of spectral indices or
that there is an additional contribution by sources which are harder
and rarer than the QSOs we have modelled here.  Additional physical
effects such as radiative transfer, the lifetime of the sources and
the opening angle for the ionizing emission may also play a role in
increasing the scatter.  Finally, systematic biases in the line
fitting procedure of Z04b may artificially increase the observed
scatter by around $0.5$ dex (\citealt{FechnerReimers04}).  As already
mentioned the observed values of $\eta<10$ and $\eta>400$ should be
considered with some caution and are probably not real
(\citealt{Shull04}).  The right column shows the result for our model
of \HeII ionizing photon emission by hot gas.  In this case the
scatter is as small as for the uniform background model. Such a small
scatter for a UVB produced by hot gas is not surprising; the number of
sources contributing to the \HeII ionizing UVB is more than an order
of magnitude larger compared to the QSO emission model.  As found for
the spatially uniform UVB, the small scatter in the column density
ratio is not consistent with the observed scatter.  M04 claim that the
contribution of hot gas to the \HeII ionizing background should
increase strongly with increasing redshift. This result appears to be
at odds with the increasing fluctuations in the \HeII opacity towards
higher redshift, suggesting that emission by hot gas actually
contributes little to the \HeII ionization rate and that the estimates
of the bremsstrahlung volume emissivity by M04 may have been too
large. This is perhaps also not too surprising as the assumed density
profile of the emitting gas in dark matter haloes is very uncertain.
More quantitative statements will require detailed modelling of the
UVB using radiative transfer which is beyond the scope of this paper.

As an additional check on our assumptions for the \HI and \HeII
effective optical depth, we plot as a dashed line in
figure~\ref{fig:eta} the median $\eta$ obtained from the artificial
spectra with a uniform UVB (second panel).  The same line is repeated
in the panels for the artificial spectra constructed using a
fluctuating UVB and the observations.  Although the scatter in the
column density ratio is much smaller for the artificial spectra with
the uniform UVB, the median value for $\eta$ is consistent with that
observed.  This is encouraging, as it suggests that our fit for the
\HeII effective optical depth at $2<z<3$ provides a reasonable
representation of the true \HeII \Lya forest opacity.  Note the
increase in the median column density ratio with redshift; a softer
UVB is required to reproduce the observed \HI and \HeII \Lya forest
opacity at higher redshift.

Figure~\ref{fig:scatter} shows how $\eta$ varies along the
line-of-sight and gives an impression of the physical scales over
which the fluctuations are correlated. The Z04b data in the range
$2.3<z<2.5$ are plotted with open diamonds in both panels.  For
comparison, results for our simulated spectra are shown as stars.  The
left panel shows the simulated data assuming a spatially uniform
background while the right panel is for our model of the UVB
fluctuations due to QSOs.  The redshifts of the simulated spectra have
been shifted to cover the same redshift range as the observational
data.  However, the actual physical distance between each data point
for the simulated spectra has been preserved to enable a fair
comparison between the scales over which $\eta$ varies.  For the
simulated spectra with a uniform UVB the data points strongly cluster
around $\eta \simeq 70$, in strong disagreement with the distribution
for the observational data. For the simulated spectra created with the
fluctuating UVB due to QSOs the spread in $\eta$ reproduces the
observational result remarkably well, although a full treatment of
radiative transfer effects may be required to produce UVB fluctuations
on the smallest physical scales (\citealt{Shull04,Maselli05}).  Note
again that it is not clear whether the underestimate of the number of
points which see a very hard spectrum, $\eta<10$ in the simulations is
real.

\begin{figure}
\begin{center}
  \includegraphics[width=0.5\textwidth]{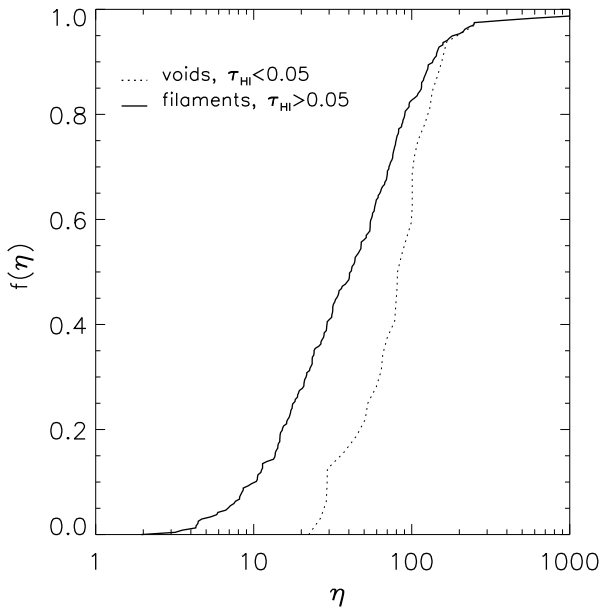}
\caption{
  The cumulative probability distribution function (CPDF) for $\eta$
  at $z=2.4$; a fluctuating UVB at the \HeII photoelectric edge due to
  QSOs has been assumed.  The solid curve shows the CPDF for \HI/\HeII
  line pairs with $\tau_{\rm HI} > 0.05$.  The dotted curve shows the
  CPDF for underdense regions ($\tau_{\rm HI} < 0.05$).  Low density
  regions in the IGM appear to see a softer radiation field,
  consistent with observational data ({\it e.g.}  \citealt{Reimers04,Shull04}).
}
\label{fig:voids}
\end{center}
\end{figure}

\subsection{The anti-correlation of the column density ratio with \HI density}

The studies of UVB fluctuations by \cite{Reimers04} and \cite{Shull04}
have also uncovered an anti-correlation between the column density
ratio $\eta$ and the density of the \HI \Lya absorbers. Regions of
high opacity within the \HI \Lya forest, corresponding to the
overdense knots, filaments and sheets which form as a natural
consequence of structure formation, tend on average to see a somewhat
harder radiation field ($\eta<100$).  In contrast, the voids in the
\HI forest account for the majority of the larger values of $\eta$,
corresponding to a softer metagalactic radiation field ($\eta>100$).
In particular, \cite{Shull04} find up to $80\%$ of the void
($\tau_{\rm HI} <0.05$) path length has $\eta>100$, as opposed to
around $40\%$ for higher density regions ($\tau_{\rm HI} > 0.05$).

Figure~\ref{fig:voids} shows a plot of the cumulative probability
distribution function (CPDF) for $\eta$ at $z=2.4$, similar to figure
$5$ in \cite{Shull04}. A fluctuating UVB at the \HeII photoelectric
edge due to QSOs was assumed.  The solid curve is the CPDF for all
identified line pairs where $\tau_{\rm H1} >0.05$. The dotted curve
shows the CPDF for underdense regions, $\tau_{\rm HI} <0.05$.  There
is very good agreement with figure $5$ in \cite{Shull04}; the
underdense regions clearly see a softer radiation field on average.
In our model the origin of this anti-correlation can be easily
understood from the spatial distribution of the QSOs which are
responsible for the radiation at $4$ Ryd. The underdense regions are
typically further away from the massive haloes which host these QSOs.
The flux at $4$ Ryd is therefore on average smaller in underdense
regions and thus the value of $\eta$ is larger.  Note, however, that
systematic uncertainties from the line fitting method will also
contribute to the the observed anti-correlation
(\citealt{FechnerReimers04}).

\section{The spectral shape of the UV background} \label{sec:UVB}
\subsection{The average softness parameter and its evolution with redshift}

We obtain an estimate of the average softness parameter from our
simulations with a spatially uniform UVB by adjusting the \HI and
\HeII ionization rates such that the artificial spectra reproduce the
observed \HI and \HeII \Lya forest opacity.  The result is shown as
the solid curve in figure~\ref{fig:softness}, corrected for numerical
resolution, with the systematic uncertainties shown by the grey
region.  We also show the same result estimated from our artificial
spectra constructed assuming a fluctuating UVB at the \HeII
photoelectric edge due to QSO emission, shown as the dot-dashed line
in figure~\ref{fig:softness}.  There is some mild evidence for an
increase of the softness parameter with increasing redshift.  We now
proceed to discuss this measurement and its uncertainties in more
detail.

\subsection{Systematic uncertainties in the estimate of the average softness parameter}

The value of $S$ we infer from our artificial spectra will depend on
our assumptions for the various cosmological and astrophysical
parameters which alter the opacity of the \Lya forest, as well as the
observational values for the \HI and \HeII effective optical depths we
adopt.  However, as we are interested in the ratio of the \HI to \HeII
metagalactic ionization rates, uncertainties in parameters on which
the \HI and \HeII opacity have the same dependence will not contribute
significantly to the overall uncertainty in $S$.  In this instance,
uncertainties from assumptions for \OBh and the temperature of the IGM
can be ignored; \OBh determines the total baryonic content, and the
\HI and \HeII recombination coefficients have the same temperature
dependence.  Consequently, altering either of these parameters will
have the same effect on the \HI and \HeII ionization rates we infer.

In contrast, the matter density as fraction of the critical density,
$\Omega_{\rm m}$, and the mass fluctuation amplitude within $8h^{-1}$
Mpc spheres, $\sigma_{8}$, both alter the rms fluctuation amplitude at
the Jeans scales, changing the gas distribution (\citealt{Bolton05}).
As already discussed, \HeII is optically thick down to lower densities
compared to H~$\rm \scriptstyle I$.  For larger \OM or $\sigma_{8}$,
the \HeII opacity decreases more relative to the \HI opacity as
absorbing gas is transferred to higher density regions.  The opposite
is also true for a smaller value of \OM or $\sigma_{8}$.  To estimate
the uncertainty on the value of $S$ in the redshift range $2<z<3$ from
these parameters, we adopt fiducial values of $\Omega_{\rm
  m}=0.26\pm0.04$ and $\sigma_{8}=0.9\pm0.1$, and run four extra
simulations to determine $S$ using parameter values at the upper and
lower end of the uncertainties.

Assumptions for the slope $\gamma$ of the power-law effective equation
of state for the low density gas (\citealt{HuiGnedin97}) will also
alter $S$.  For a flatter effective equation of state, a larger
fraction of the optically thick \HeII gas will have a higher
temperature compared to the \HI gas, reducing the inferred \HeII
ionization rate to a greater extent than the \HI ionization rate,
increasing the inferred softness parameter.  The converse is also
true; a larger value of $\gamma$ will increase the \HeII opacity
substantially more than that for H~$\rm \scriptstyle I$, lowering the
inferred $S$.  The plausible range for $\gamma$ is $1.0<\gamma<1.6$
(\citealt{HuiGnedin97,Schaye00}).  We estimate the effect of $\gamma$
on $S$ by artificially rescaling the effective equation of state of
$15-200$ simulation data by pivoting the temperature-density relation
around the mean gas density.

\begin{table} 
  \centering
  \caption{
    Percentage error budget for the softness parameter at
  $z=[2.1,2.4,2.8]$ from estimates of various cosmological and
  astrophysical parameters, listed approximately in order of
  importance.  The total error is obtained by adding the individual
  errors in quadrature.  Our estimate for the softness parameter at 
  these redshifts and the corresponding uncertainty is listed in the last row.
  A spatially uniform UVB has been assumed.}

  \begin{tabular}{c|c|c|c}
\hline
Parameter & $z=2.1$ & $z=2.4$ & $z=2.8$ \\  
& & &\\
$\tau_{\rm HeII}$  & $^{+61}_{-41}$ per cent & $^{+79}_{-42}$ per cent & $^{+188}_{-41}$ per cent  \\
& & & \\
$\tau_{\rm HI}$   & $^{+24}_{-18}$ per cent & $^{+22}_{-17}$ per cent & $^{+25}_{-18}$ per cent  \\
& & & \\
$\gamma$        & $^{+25}_{-10}$ per cent & $^{+23}_{-13}$ per cent & $^{+20}_{-17}$ per cent  \\
& & & \\
Numerical         &  $\pm$10 per cent &  $\pm$10 per cent  &  $\pm$10 per cent  \\
& & & \\
$\sigma_{8}$     & $^{+6}_{-8}$ per cent & $\pm$9 per cent & $\pm$10 per cent  \\  
& & & \\
$\Omega_{\rm m}$  & $^{+5}_{-6}$ per cent & $\pm$6 per cent & $^{+10}_{-7}$ per cent  \\
& & &\\
Total & $^{+72}_{-48}$ per cent & $^{+87}_{-49}$ per cent & $^{+191}_{-50}$ per cent \\
& & &  \\
S & 139$^{+99}_{-67}$ & 196$^{+170}_{-97}$ & 301$^{+576}_{-151}$ \\
\hline  
\label{tab:errors}
\end{tabular}
\end{table}

Finally, the uncertainty in the adopted values for $\tau_{\rm HI}$ and
$\tau_{\rm HeII}$ must be considered.  We estimate the uncertainty in
$\tau_{\rm HI}$ by binning the 1$\sigma$ measurement errors of
\cite{Schaye03} as described in section $2.3$.  We use the error bars
shown in figure~\ref{fig:tau} to estimate the uncertainty in our
assumption for the \HeII effective optical depth.  These have been
computed using the fluctuating UVB model due to QSOs described in
section~\ref{sec:toymodel}; the lower and upper error bars correspond
to the $25^{\rm th}$ and $75^{\rm th}$ percentiles of $\tau_{\rm
  HeII}$ taken from each set of $1024$ spectra at the relevant
redshift.  An additional uncertainty of $10$ per cent is added to take
into account the marginal convergence of the simulations.  The final
error budget for $S$ at $z=[2.1,2.4,2.8]$ is summarised in
Table~\ref{tab:errors}.  The dominant uncertainty is the measurement
of the \HeII effective optical depth.

\subsection{Comparison to the UVB model of Madau et al. and other
  observational estimates}

\begin{figure*}
  \centering
  \begin{minipage}{180mm}
    \begin{center}
      \psfig{figure=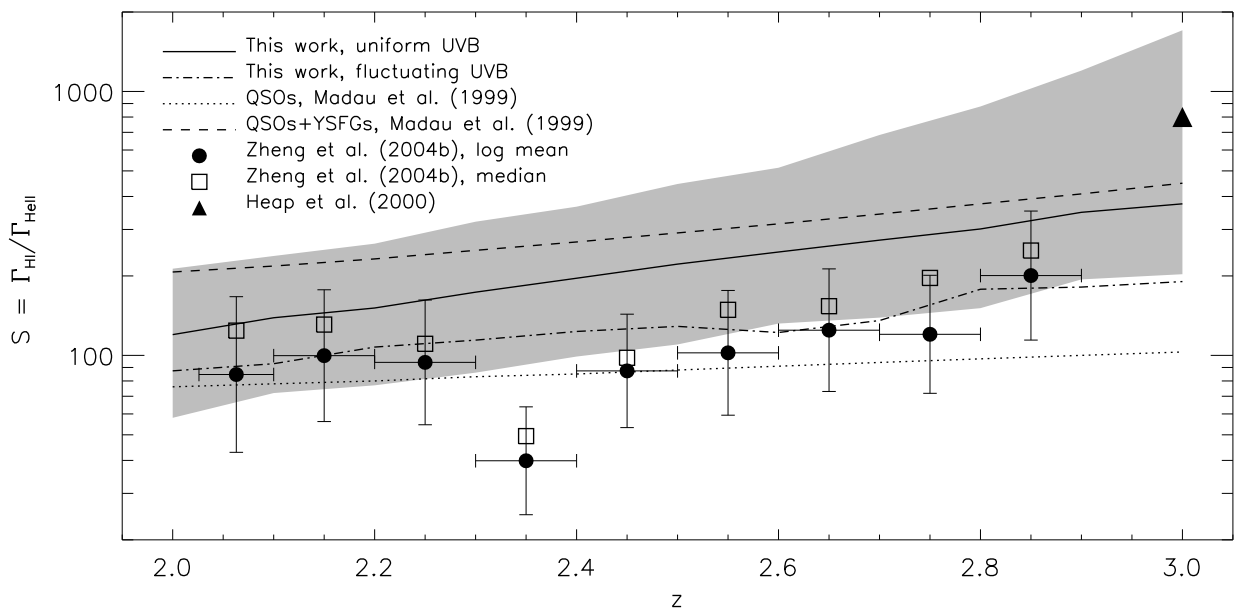,width=1.0\textwidth}
      
      \caption{
        Our determination of the average softness parameter of the UVB
        compared to the model of MHR99 and other observational
        estimates. The solid curve shows our results $S$ corrected for
        resolution effects assuming a spatially uniform UVB.  The grey
        shaded area indicates the theoretical uncertainty.  The
        dot-dashed curves shows our result for the case of a spatially
        fluctuating \HeII ionization rate.  The dotted and dashed
        curves represent the value of $S$ predicted by the updated UVB
        models of MHR99, based respectively on contributions from QSOs
        alone and QSOs plus YSFGs.  The filled circles with error bars
        show the results of Z04b calculated from the logarithmic mean
        of the measured column density ratios $\eta$ using
        equation~\ref{eq:soft}.  The open squares show the median
        value of $S$ in each redshift bin for the same data. Finally,
        the filled triangle represents the determination of $S$ at
        $z\simeq 3$ by Heap \etal (2000).}
      \label{fig:softness}
    \end{center}
  \end{minipage}
\end{figure*}

In figure~\ref{fig:softness} we compare the results of our analysis of
the average softness parameter inferred from observed \HI and \HeII
effective optical depths with the UVB model of Madau et al. (1999)
(hereinafter MHR99) and other observational estimates of $S$.  The
dotted and dashed curves show the softness parameter of the updated
UVB models of MHR99 (see \citealt{Bolton05} for a detailed
description) for QSOs alone and QSOs plus young star forming galaxies
(YSFGs).  Typically $S \leq 100$ for QSO dominated UVB models
(\citealt{HaardtMadau96,Fardal98}), although if there is a wide range
in QSO spectral indices a QSO dominated UVB could be as soft as
$S\simeq 400$ (\citealt{Telfer02}).  Due to their softer spectra,
YSFGs only provide a significant boost to the UVB amplitude at the \HI
photoelectric edge, so that typically one expects $S \geq 200$ for a
UVB with contributions from QSOs and YSFGs.

Our results for the spatially uniform UVB model agree very well with
the softness parameter of the updated MHR99 model with contributions
from QSOs and YSFGs.  At the lowest redshift $z\sim 2$ there is some
indication that the MHR99 model may slightly overpredict the softness
parameter. Note that the MHR99 UVB model also assumes a spatially
uniform background, allowing a fair comparison.  Our analysis thus
adds to the growing evidence that the amplitude of the UVB at the \HI
photoelectric edge cannot be accounted for by QSO emissivities alone;
a substantial contribution from YSFGs seems to be required ({\it e.g.}
\citealt{GirouxShull97,Zhang97,Theuns98,Bianchi01,Kriss01,Haehnelt01,
  Smette02,Zheng04b,Shull04,Aguirre04,Bolton05,Kirkman05}).  Note,
however, that as discussed before, UV fluctuations will play a
significant role in altering $S$ on local scales, especially at
$z>2.8$.  Some regions of low opacity are obviously subject to a
harder ionizing radiation field, producing values of $S$ outside our
error range.  Similarly, overdense regions which are optically thick
to \HeII ionizing photons can be subject to a softer radiation field.

The spatial fluctuations also affect the average \HeII ionization rate
and thus the inferred average softness parameter. The dot-dashed curve
in figure~\ref{fig:softness} shows the average softness parameter
obtained from our model of a spatially fluctuating UVB due to QSO
emission.  The average softness parameter inferred for the spatially
fluctuating \HeII ionization rate is systematically smaller ($0.15$
dex) than that obtained for a spatially uniform UVB.  This is because
in regions close to QSOs, corresponding to voids in the \HeII \Lya
forest, raising the \HeII ionization rate further will not decrease
the effective optical depth to a great extent once the local optical
depth is already low.  The average \HeII ionization rate will thus be
biased low if a uniform background is assumed.  This effect is also
found for UVB fluctuations at the hydrogen ionization edge ({\it e.g.}
\citealt{GnedinHamilton02,MeiksinWhite04}).

We also plot the data of Z04b and \cite{Heap00} in
figure~\ref{fig:softness}.  The logarithmic mean of the softness
parameter calculated from the Z04b column density measurements is shown
by the filled circles with error bars.  The median value in each bin
is indicated with an open square.  We have only used data where both
the \HI and \HeII components have measured column densities; the
analysis of our artificial spectra using Voigt profile fitting
indicates that these measurements should recover the UVB spectral
shape well for the line-of-sight considered, with the median providing
a better estimate of $S$. The results of Z04b agree very well with our
analysis for a fluctuating \HeII ionization rate.  Note that the
results of Z04b cannot be compared directly to the predictions of the
MHR99 model which assumes a spatially uniform UVB.

\section{Conclusions}

We have used state-of the art hydrodynamical simulations to obtain
improved measurements of the spectral shape of the ionizing UV
background from the \HI and \HeII \Lya forest and to investigate the
origin of the large spatial fluctuations observed in the \HeII to \HI
column density ratio.

Using artificial absorption spectra we have shown that the softness
parameter of the UV background can be accurately inferred from the
column density ratio of \HeII and \HI absorption lines obtained by
applying standard Voigt profile fitting routines. Uncertainties in the
identification of corresponding \HI and \HeII absorption and errors in
the determination of the column densities contribute little to the
large fluctuations observed in the \HeII and \HI column density ratio,
although one should be cautious of extreme values of $\eta$.  These
fluctuations must be due to genuine spatial variations in the spectral
shape of the UVB.  A model where the \HeII ionization rate fluctuates
due to variation in the number, luminosity and spectral shape of a
small number of QSOs reproduces the observed spatial variations of the
\HeII and \HI column density ratio, as well as the observed
anti-correlation of the column density ratio with the \HI density.
This is in good agreement with previous suggestions that the observed
\HeII \Lya forest spectra at $z\sim 2-3.5$ probes the tail end of the
reionization of \HeII by QSOs.  The large fluctuations observed in the
\HeII and \HI column density ratio argue strongly against a
significant contribution of emission by hot gas to the \HeII
ionization rate at $2<z<3$.

We further obtain new constraints and error estimates for the mean
softness parameter of the UVB and its evolution using the observed
effective optical depth of the \HeII and \HI \Lya forest.  We find
that the inferred softness parameter is consistent with a UVB with
contributions from QSOs and YSFGs within our uncertainties.  The
dominant contribution to the uncertainty on $S$ comes from the \HeII
effective optical depth.  Our results for the mean
softness parameter add to the growing body of evidence suggesting that
a UVB dominated by emission from QSOs alone cannot reproduce the
observed effective optical depth of the \Lya forest in QSO spectra.

\section*{Acknowledgements}
We thank Volker Springel for his advice and for making GADGET-2
available.  We are also grateful to Francesco Haardt for making his
updated UV background model available to us.  JSB thanks John Regan
for assistance with running simulations, and Natasha Maddox and Paul
Hewett for helpful discussions.  This research was conducted in
cooperation with SGI/Intel utilising the Altix 3700 supercomputer
COSMOS at the Department of Applied Mathematics and Theoretical
Physics in Cambridge.  COSMOS is a UK-CCC facility which is supported
by HEFCE and PPARC.  We also acknowledge support from the European
Community Research and Training Network 'The Physics of the
Intergalactic Medium'.  JSB, MGH, MV and RFC thank PPARC for financial
support.  We also thank the referee, Mike Shull, for a very detailed
and helpful report.

\label{lastpage}


\begin{thebibliography}{99}

\bibitem[\protect\citeauthoryear{{Abel}, {Anninos}, {Zhang} \& {Norman}}{{Abel}
  et~al.}{1997}]{Abel97}
{Abel} T.,  {Anninos} P.,  {Zhang} Y.,    {Norman} M.~L.,  1997, NewA,
  2, 181

\bibitem[\protect\citeauthoryear{{Abel} \& {Haehnelt}}{{Abel} \&
  {Haehnelt}}{1999}]{AbelHaehnelt99}
{Abel} T.,  {Haehnelt} M.~G.,  1999, ApJ, 520, L13

\bibitem[\protect\citeauthoryear{{Aguirre}, {Schaye}, {Kim}, {Theuns}, {Rauch}
  \& {Sargent}}{{Aguirre} et~al.}{2004}]{Aguirre04}
{Aguirre} A.,  {Schaye} J.,  {Kim} T.,  {Theuns} T.,  {Rauch} M.,    {Sargent}
  W.~L.~W.,  2004, ApJ, 602, 38

\bibitem[\protect\citeauthoryear{{Anderson}, {Hogan}, {Williams} \&
  {Carswell}}{{Anderson} et~al.}{1999}]{Anderson99}
{Anderson} S.~F.,  {Hogan} C.~J.,  {Williams} B.~F.,    {Carswell} R.~F.,
  1999, AJ, 117, 56

\bibitem[\protect\citeauthoryear{{Anninos}, {Zhang}, {Abel} \&
  {Norman}}{{Anninos} et~al.}{1997}]{Anninos97}
{Anninos} P.,  {Zhang} Y.,  {Abel} T.,    {Norman} M.~L.,  1997, NewA,
  2, 209

\bibitem[\protect\citeauthoryear{{Bernardi} et al.}{{Bernardi} 
    et~al.}{2003}]{Bernardi03}
{Bernardi} M., et al. 2003, AJ, 125, 32

\bibitem[\protect\citeauthoryear{{Bi} \& {Davidsen}}{{Bi} \&
  {Davidsen}}{1997}]{Bi97}
{Bi} H.,  {Davidsen} A.~F.,  1997, ApJ, 479, 523

\bibitem[\protect\citeauthoryear{{Bi}, {Boerner} \& {Chu}}{{Bi}
  et~al.}{1992}]{Bi92}
{Bi} H.~G.,  {Boerner} G.,    {Chu} Y.,  1992, A\&A, 266, 1

\bibitem[\protect\citeauthoryear{{Bianchi}, {Cristiani} \& {Kim}}{{Bianchi}
  et~al.}{2001}]{Bianchi01}
{Bianchi} S.,  {Cristiani} S.,    {Kim} T.-S.,  2001, A\&A, 376, 1

\bibitem[\protect\citeauthoryear{{Bolton}, {Meiksin} \& {White}}{{Bolton}
  et~al.}{2004}]{Bolton04}
{Bolton} J.,  {Meiksin} A.,    {White} M.,  2004, MNRAS, 348, L43

\bibitem[\protect\citeauthoryear{{Bolton}, {Haehnelt}, {Viel} \&
  {Springel}}{{Bolton} et~al.}{2005}]{Bolton05}
{Bolton} J.~S.,  {Haehnelt} M.~G.,  {Viel} M.,    {Springel} V.,  2005, MNRAS,
  357, 1178

\bibitem[\protect\citeauthoryear{{Boyle}, {Shanks}, {Croom}, {Smith}, {Miller},
  {Loaring} \& {Heymans}}{{Boyle} et~al.}{2000}]{Boyle00}
{Boyle} B.~J.,  {Shanks} T.,  {Croom} S.~M.,  {Smith} R.~J.,  {Miller} L.,
  {Loaring} N.,    {Heymans} C.,  2000, MNRAS, 317, 1014

\bibitem[\protect\citeauthoryear{{Cen}}{{Cen}}{1992}]{Cen92}
{Cen} R.,  1992, ApJS, 78, 341

\bibitem[\protect\citeauthoryear{{Croft}}{{Croft}}{2004}]{Croft04}
{Croft} R.~A.~C.,  2004, ApJ, 610, 642

\bibitem[\protect\citeauthoryear{{Croft}, {Weinberg}, {Katz} \&
  {Hernquist}}{{Croft} et~al.}{1997}]{Croft97}
{Croft} R.~A.~C.,  {Weinberg} D.~H.,  {Katz} N.,    {Hernquist} L.,  1997,
  ApJ, 488, 532

\bibitem[\protect\citeauthoryear{{Croft}, {Weinberg}, {Pettini}, {Hernquist} \&
  {Katz}}{{Croft} et~al.}{1999}]{Croft99}
{Croft} R.~A.~C.,  {Weinberg} D.~H.,  {Pettini} M.,  {Hernquist} L.,    {Katz}
  N.,  1999, ApJ, 520, 1

\bibitem[\protect\citeauthoryear{{Croom}, {Smith}, {Boyle}, {Shanks}, {Miller},
  {Outram} \& {Loaring}}{{Croom} et~al.}{2004}]{Croom04}
{Croom} S.~M.,  {Smith} R.~J.,  {Boyle} B.~J.,  {Shanks} T.,  {Miller} L.,
  {Outram} P.~J.,    {Loaring} N.~S.,  2004, MNRAS, 349, 1397

\bibitem[\protect\citeauthoryear{{Davidsen}, {Kriss} \& {Wei}}{{Davidsen}
  et~al.}{1996}]{Davidsen96}
{Davidsen} A.~F.,  {Kriss} G.~A.,    {Wei} Z.,  1996, Nature, 380, 47

\bibitem[\protect\citeauthoryear{{Fardal}, {Giroux} \& {Shull}}{{Fardal}
  et~al.}{1998}]{Fardal98}
{Fardal} M.~A.,  {Giroux} M.~L.,    {Shull} J.~M.,  1998, AJ, 115, 2206

\bibitem[\protect\citeauthoryear{{Fechner} \& {Reimers}}{{Fechner} \&
  {Reimers}}{2004}]{FechnerReimers04}
{Fechner} C.,  {Reimers} D.,  2004, preprint (astro-ph/0410622)

\bibitem[\protect\citeauthoryear{{Ferrarese} \& {Merritt}}{{Ferrarese} \&
  {Merritt}}{2000}]{FerrareseMerritt00}
{Ferrarese} L.,  {Merritt} D.,  2000, ApJ, 539, L9

\bibitem[\protect\citeauthoryear{{Giroux} \& {Shull}}{{Giroux} \&
  {Shull}}{1997}]{GirouxShull97}
{Giroux} M.~L.,  {Shull} J.~M.,  1997, AJ, 113, 1505

\bibitem[\protect\citeauthoryear{{Gleser}, {Nusser}, {Benson}, {Ohno} \&
  {Sugiyama}}{{Gleser} et~al.}{2005}]{Gleser04}
{Gleser} L.,  {Nusser} A.,  {Benson} A.~J.,  {Ohno} H.,    {Sugiyama} N.,
  2005, MNRAS, 361, 1399

\bibitem[\protect\citeauthoryear{{Gnedin} \& {Hamilton}}{{Gnedin} \&
  {Hamilton}}{2002}]{GnedinHamilton02}
{Gnedin} N.~Y.,  {Hamilton} A.~J.~S.,  2002, MNRAS, 334, 107

\bibitem[\protect\citeauthoryear{{Haardt} \& {Madau}}{{Haardt} \&
  {Madau}}{1996}]{HaardtMadau96}
{Haardt} F.,  {Madau} P.,  1996, ApJ, 461, 20

\bibitem[\protect\citeauthoryear{{Haehnelt}, {Madau}, {Kudritzki} \&
  {Haardt}}{{Haehnelt} et~al.}{2001}]{Haehnelt01}
{Haehnelt} M.~G.,  {Madau} P.,  {Kudritzki} R.,    {Haardt} F.,  2001, ApJ,
  549, L151

\bibitem[\protect\citeauthoryear{{Heap}, {Williger}, {Smette}, {Hubeny},
  {Sahu}, {Jenkins}, {Tripp} \& {Winkler}}{{Heap} et~al.}{2000}]{Heap00}
{Heap} S.~R.,  {Williger} G.~M.,  {Smette} A.,  {Hubeny} I.,  {Sahu} M.~S.,
  {Jenkins} E.~B.,  {Tripp} T.~M.,    {Winkler} J.~N.,  2000, ApJ, 534, 69

\bibitem[\protect\citeauthoryear{{Hernquist}, {Katz}, {Weinberg} \&
  {Miralda-Escud{\' e}}}{{Hernquist} et~al.}{1996}]{Hernquist96}
{Hernquist} L.,  {Katz} N.,  {Weinberg} D.~H.,    {Miralda-Escud{\' e}} J.,
  1996, ApJ, 457, L51

\bibitem[\protect\citeauthoryear{{Hogan}, {Anderson} \& {Rugers}}{{Hogan}
  et~al.}{1997}]{Hogan97}
{Hogan} C.~J.,  {Anderson} S.~F.,    {Rugers} M.~H.,  1997, AJ, 113, 1495

\bibitem[\protect\citeauthoryear{{Hui} \& {Gnedin}}{{Hui} \&
  {Gnedin}}{1997}]{HuiGnedin97}
{Hui} L.,  {Gnedin} N.~Y.,  1997, MNRAS, 292, 27

\bibitem[\protect\citeauthoryear{{Jakobsen}}{{Jakobsen}}{1996}]{Jakobsen96}
{Jakobsen} P.,  1996, in Benvenuti P., Macchetto F.D., Schreier E.J., eds,
  Science with the Hubble Space Telescope - II. Space Telescope Science
  Institute, Baltimore, p.153

\bibitem[\protect\citeauthoryear{{Jakobsen}, {Boksenberg}, {Deharveng},
  {Greenfield}, {Jedrzejewski} \& {Paresce}}{{Jakobsen}
  et~al.}{1994}]{Jakobsen94}
{Jakobsen} P.,  {Boksenberg} A.,  {Deharveng} J.~M.,  {Greenfield} P.,
  {Jedrzejewski} R.,    {Paresce} F.,  1994, Nature, 370, 35

\bibitem[\protect\citeauthoryear{{Jakobsen}, {Jansen}, {Wagner} \&
  {Reimers}}{{Jakobsen} et~al.}{2003}]{Jakobsen03}
{Jakobsen} P.,  {Jansen} R.~A.,  {Wagner} S.,    {Reimers} D.,  2003, A\&A,
  397, 891

\bibitem[\protect\citeauthoryear{{Katz}, {Weinberg} \& {Hernquist}}{{Katz}
  et~al.}{1996}]{Katz96}
{Katz} N.,  {Weinberg} D.~H.,    {Hernquist} L.,  1996, ApJS, 105, 19

\bibitem[\protect\citeauthoryear{{Kirkman} et al.}{{Kirkman} 
    et~al.}{2005}]{Kirkman05}
{Kirkman} D.,  et al. 2005, MNRAS, 360, 1373

\bibitem[\protect\citeauthoryear{{Kriss} et al.}{{Kriss} 
    et~al.}{2001}]{Kriss01}
{Kriss} G.~A.,  et al. 2001, Science, 293, 1112

\bibitem[\protect\citeauthoryear{{Madau}, {Haardt} \& {Rees}}{{Madau}
  et~al.}{1999}]{Madau99}
{Madau} P.,  {Haardt} F.,    {Rees} M.~J.,  1999, ApJ, 514, 648

\bibitem[\protect\citeauthoryear{{Madau} \& {Meiksin}}{{Madau} \&
  {Meiksin}}{1994}]{MadauMeiksin94}
{Madau} P.,  {Meiksin} A.,  1994, ApJ, 433, L53

\bibitem[\protect\citeauthoryear{{Maselli} \& {Ferrara}}{{Maselli} \&
  {Ferrara}}{2005}]{Maselli05}
{Maselli} A.,  {Ferrara} A.,  2005, preprint (astro-ph/0510258)

\bibitem[\protect\citeauthoryear{{McDonald} \& {Miralda-Escud{\'
  e}}}{{McDonald} \& {Miralda-Escud{\' e}}}{2001}]{McDonaldMiraldaEscude01}
{McDonald} P.,  {Miralda-Escud{\' e}} J.,  2001, ApJ, 549, L11

\bibitem[\protect\citeauthoryear{{McDonald}, {Miralda-Escud{\' e}}, {Rauch},
  {Sargent}, {Barlow} \& {Cen}}{{McDonald} et~al.}{2001}]{McDonald01}
{McDonald} P.,  {Miralda-Escud{\' e}} J.,  {Rauch} M.,  {Sargent} W.~L.~W.,
  {Barlow} T.~A.,    {Cen} R.,  2001, ApJ, 562, 52

\bibitem[\protect\citeauthoryear{{Meiksin} \& {White}}{{Meiksin} \&
  {White}}{2004}]{MeiksinWhite04}
{Meiksin} A.,  {White} M.,  2004, MNRAS, 350, 1107

\bibitem[\protect\citeauthoryear{{Miniati}, {Ferrara}, {White} \&
  {Bianchi}}{{Miniati} et~al.}{2004}]{Miniati04}
{Miniati} F.,  {Ferrara} A.,  {White} S.~D.~M.,    {Bianchi} S.,  2004, MNRAS,
  348, 964 (M04)

\bibitem[\protect\citeauthoryear{{Miralda-Escud{\' e}}}{{Miralda-Escud{\'
  e}}}{1993}]{MiraldaEscude93}
{Miralda-Escud{\' e}} J.,  1993, MNRAS, 262, 273

\bibitem[\protect\citeauthoryear{{Miralda-Escud{\' e}}, {Cen}, {Ostriker} \&
  {Rauch}}{{Miralda-Escud{\' e}} et~al.}{1996}]{MiraldaEscude96}
{Miralda-Escud{\' e}} J.,  {Cen} R.,  {Ostriker} J.~P.,    {Rauch} M.,  1996,
  ApJ, 471, 582

\bibitem[\protect\citeauthoryear{{Miralda-Escud{\' e}}, {Haehnelt} \&
  {Rees}}{{Miralda-Escud{\' e}} et~al.}{2000}]{MiraldaEscude00}
{Miralda-Escud{\' e}} J.,  {Haehnelt} M.,    {Rees} M.~J.,  2000, ApJ, 530, 1

\bibitem[\protect\citeauthoryear{{Miralda-Escud{\' e}} \&
  {Ostriker}}{{Miralda-Escud{\' e}} \&
  {Ostriker}}{1990}]{MiraldaEscudeOstriker90}
{Miralda-Escud{\' e}} J.,  {Ostriker} J.~P.,  1990, ApJ, 350, 1

\bibitem[\protect\citeauthoryear{{Rauch}}{{Rauch}}{1998}]{Rauch98}
{Rauch} M.,  1998, ARA\&A, 36, 267

\bibitem[\protect\citeauthoryear{{Rauch} et al.}{{Rauch} 
    et~al.}{1997}]{Rauch97}
{Rauch} M.,  et al. 1997, ApJ, 489, 7

\bibitem[\protect\citeauthoryear{{Reimers}, {Fechner}, {Hagen}, {Jakobsen},
  {Tytler} \& {Kirkman}}{{Reimers} et~al.}{2005a}]{Reimers05a}
{Reimers} D.,  {Fechner} C.,  {Hagen} H.-J.,  {Jakobsen} P.,  {Tytler} D.,
  {Kirkman} D.,  2005a, A\&A, 442, 63

\bibitem[\protect\citeauthoryear{{Reimers}, {Fechner}, {Kriss}, {Shull},
  {Baade}, {Moos}, {Songaila} \& {Simcoe}}{{Reimers} et~al.}{2004}]{Reimers04}
{Reimers} D.,  {Fechner} C.,  {Kriss} G.,  {Shull} M.,  {Baade} R.,  {Moos} W.,
   {Songaila} A.,    {Simcoe} R.,  2004, preprint (astro-ph/0410588)

\bibitem[\protect\citeauthoryear{{Reimers}, {Hagen}, {Schramm}, {Kriss} \&
  {Shull}}{{Reimers} et~al.}{2005b}]{Reimers05b}
{Reimers} D.,  {Hagen} H.-J.,  {Schramm} J.,  {Kriss} G.~A.,    {Shull} J.~M.,
  2005b, A\&A, 436, 465

\bibitem[\protect\citeauthoryear{{Reimers}, {Kohler}, {Wisotzki}, {Groote},
  {Rodriguez-Pascual} \& {Wamsteker}}{{Reimers} et~al.}{1997}]{Reimers97}
{Reimers} D.,  {Kohler} S.,  {Wisotzki} L.,  {Groote} D.,  {Rodriguez-Pascual}
  P.,    {Wamsteker} W.,  1997, A\&A, 327, 890

\bibitem[\protect\citeauthoryear{{Richards} et al.}{{Richards} 
    et~al.}{2005}]{Richards05}
{Richards} G.~T.,  et al. 2005, MNRAS, 360, 839

\bibitem[\protect\citeauthoryear{{Ricotti}, {Gnedin} \& {Shull}}{{Ricotti}
  et~al.}{2000}]{Ricotti00}
{Ricotti} M.,  {Gnedin} N.~Y.,    {Shull} J.~M.,  2000, ApJ, 534, 41

\bibitem[\protect\citeauthoryear{{Schaye}, {Aguirre}, {Kim}, {Theuns}, {Rauch}
  \& {Sargent}}{{Schaye} et~al.}{2003}]{Schaye03}
{Schaye} J.,  {Aguirre} A.,  {Kim} T.,  {Theuns} T.,  {Rauch} M.,    {Sargent}
  W.~L.~W.,  2003, ApJ, 596, 768

\bibitem[\protect\citeauthoryear{{Schaye}, {Theuns}, {Rauch}, {Efstathiou} \&
  {Sargent}}{{Schaye} et~al.}{2000}]{Schaye00}
{Schaye} J.,  {Theuns} T.,  {Rauch} M.,  {Efstathiou} G.,    {Sargent}
  W.~L.~W.,  2000, MNRAS, 318, 817

\bibitem[\protect\citeauthoryear{{Scott}, {Kriss}, {Brotherton}, {Green},
  {Hutchings}, {Shull} \& {Zheng}}{{Scott} et~al.}{2004}]{Scott04}
{Scott} J.~E.,  {Kriss} G.~A.,  {Brotherton} M.,  {Green} R.~F.,  {Hutchings}
  J.,  {Shull} J.~M.,    {Zheng} W.,  2004, ApJ, 615, 135

\bibitem[\protect\citeauthoryear{{Shull}, {Tumlinson}, {Giroux}, {Kriss} \&
  {Reimers}}{{Shull} et~al.}{2004}]{Shull04}
{Shull} J.~M.,  {Tumlinson} J.,  {Giroux} M.~L.,  {Kriss} G.~A.,    {Reimers}
  D.,  2004, ApJ, 600, 570

\bibitem[\protect\citeauthoryear{{Smette}, {Heap}, {Williger}, {Tripp},
  {Jenkins} \& {Songaila}}{{Smette} et~al.}{2002}]{Smette02}
{Smette} A.,  {Heap} S.~R.,  {Williger} G.~M.,  {Tripp} T.~M.,  {Jenkins}
  E.~B.,    {Songaila} A.,  2002, ApJ, 564, 542

\bibitem[\protect\citeauthoryear{{Sokasian}, {Abel} \& {Hernquist}}{{Sokasian}
  et~al.}{2002}]{Sokasian02}
{Sokasian} A.,  {Abel} T.,    {Hernquist} L.,  2002, MNRAS, 332, 601

\bibitem[\protect\citeauthoryear{{Songaila}}{{Songaila}}{1998}]{Songaila98}
{Songaila} A.,  1998, AJ, 115, 2184

\bibitem[\protect\citeauthoryear{{Spergel} et al.}{{Spergel} 
    et~al.}{2003}]{Spergel03}
{Spergel} D.~N.,  et al. 2003, ApJS, 148, 175

\bibitem[\protect\citeauthoryear{{Springel}}{{Springel}}{2005}]{Springel05}
{Springel} V.,  2005, preprint (astro-ph/0505010)

\bibitem[\protect\citeauthoryear{{Springel}, {Yoshida} \& {White}}{{Springel}
  et~al.}{2001}]{Springel01}
{Springel} V.,  {Yoshida} N.,    {White} S.~D.~M.,  2001, NewA, 6, 79

\bibitem[\protect\citeauthoryear{{Storrie-Lombardi}, {McMahon}, {Irwin} \&
  {Hazard}}{{Storrie-Lombardi} et~al.}{1994}]{StorrieLombardi94}
{Storrie-Lombardi} L.~J.,  {McMahon} R.~G.,  {Irwin} M.~J.,    {Hazard} C.,
  1994, ApJ, 427, L13

\bibitem[\protect\citeauthoryear{{Telfer}, {Zheng}, {Kriss} \&
  {Davidsen}}{{Telfer} et~al.}{2002}]{Telfer02}
{Telfer} R.~C.,  {Zheng} W.,  {Kriss} G.~A.,    {Davidsen} A.~F.,  2002, ApJ,
  565, 773

\bibitem[\protect\citeauthoryear{{Theuns}, {Leonard}, {Efstathiou}, {Pearce} \&
  {Thomas}}{{Theuns} et~al.}{1998}]{Theuns98}
{Theuns} T.,  {Leonard} A.,  {Efstathiou} G.,  {Pearce} F.~R.,    {Thomas}
  P.~A.,  1998, MNRAS, 301, 478

\bibitem[\protect\citeauthoryear{{Theuns}, {Schaye}, {Zaroubi}, {Kim},
  {Tzanavaris} \& {Carswell}}{{Theuns} et~al.}{2002}]{Theuns02}
{Theuns} T.,  {Schaye} J.,  {Zaroubi} S.,  {Kim} T.,  {Tzanavaris} P.,
  {Carswell} B.,  2002, ApJ, 567, L103

\bibitem[\protect\citeauthoryear{{Tytler}, {Fan}, {Burles}, {Cottrell},
  {Davis}, {Kirkman} \& {Zuo}}{{Tytler} et~al.}{1995}]{Tytler95}
{Tytler} D.,  {Fan} X.-M.,  {Burles} S.,  {Cottrell} L.,  {Davis} C.,
  {Kirkman} D.,    {Zuo} L.,  1995, in Meylan G., ed, Proc. of ESO Workshop,
  QSO Absorption Lines. Springer-Verlag, Berlin, p.289

\bibitem[\protect\citeauthoryear{{Wadsley}, {Hogan} \& {Anderson}}{{Wadsley}
  et~al.}{1999}]{Wadsley98}
{Wadsley} J.~W.,  {Hogan} C.~J.,    {Anderson} S.~F.,  1999, in Holt S.,
  Smith E., eds, AIP Conf. Proc. 470: After the Dark Ages: When Galaxies were
  Young (the Universe at 2 $<$ z $<$ 5). American Institute of Physics Press,
  p.273

\bibitem[\protect\citeauthoryear{{Walker}, {Steigman}, {Kang}, {Schramm} \&
  {Olive}}{{Walker} et~al.}{1991}]{Walker91}
{Walker} T.~P.,  {Steigman} G.,  {Kang} H.,  {Schramm} D.~M.,    {Olive} K.~A.,
   1991, ApJ, 376, 51

\bibitem[\protect\citeauthoryear{{Weinberg}, {Hernquist}, {Katz}, {Croft} \&
  {Miralda-Escud{\' e}}}{{Weinberg} et~al.}{1998}]{Weinberg98}
{Weinberg} D.~H.,  {Hernquist} L.,  {Katz} N.,  {Croft} R.,
  {Miralda-Escud{\' e}} J.,  1998, in Petitjean P., Charlot S., eds, Proc.
  13th IAP Colloq., Structure and Evolution in the Intergalactic Medium.
  Editions Fronti{\' e}res, Paris, p.133

\bibitem[\protect\citeauthoryear{{Zhang}, {Anninos}, {Norman} \&
  {Meiksin}}{{Zhang} et~al.}{1997}]{Zhang97}
{Zhang} Y.,  {Anninos} P.,  {Norman} M.~L.,    {Meiksin} A.,  1997, ApJ, 485,
  496

\bibitem[\protect\citeauthoryear{{Zhang}, {Meiksin}, {Anninos} \&
  {Norman}}{{Zhang} et~al.}{1998}]{Zhang98}
{Zhang} Y.,  {Meiksin} A.,  {Anninos} P.,    {Norman} M.~L.,  1998, ApJ, 495,
  63

\bibitem[\protect\citeauthoryear{{Zheng}, {Chiu}, {Anderson}, {Schneider},
  {Hogan}, {York}, {Burles} \& {Brinkmann}}{{Zheng} et~al.}{2004a}]{Zheng04a}
{Zheng} W.,  {Chiu} K.,  {Anderson} S.~F.,  {Schneider} D.~P.,  {Hogan} C.~J.,
  {York} D.~G.,  {Burles} S.,    {Brinkmann} J.,  2004a, AJ, 127, 656

\bibitem[\protect\citeauthoryear{{Zheng} \& {Davidsen}}{{Zheng} \&
  {Davidsen}}{1995}]{ZhengDavidsen95}
{Zheng} W.,  {Davidsen} A.,  1995, ApJ, 440, L53

\bibitem[\protect\citeauthoryear{{Zheng}, {Davidsen} \& {Kriss}}{{Zheng}
  et~al.}{1998}]{Zheng98}
{Zheng} W.,  {Davidsen} A.~F.,    {Kriss} G.~A.,  1998, AJ, 115, 391

\bibitem[\protect\citeauthoryear{{Zheng} et al.}{{Zheng} 
    et~al.}{2004b}]{Zheng04b}
{Zheng} W.,  et al. 2004b, ApJ, 605, 631 (Z04b)

\end{thebibliography}
\end{document}